\documentclass[11pt,a4paper]{article}
\pdfoutput=1

\usepackage{jheppub}
\usepackage[english]{babel}
\usepackage[utf8]{inputenc} 
\usepackage{lmodern}
\usepackage{amsmath, amsthm, amssymb}

\usepackage{tikz-feynman}
\tikzfeynmanset{compat=1.1.0}

\usepackage{caption}
\usepackage{subfig}

\usepackage{amsfonts}

\usepackage{makeidx}
\usepackage{bbm}
\usepackage{slashed}
\usepackage{booktabs}
\usepackage{braket}
\usepackage{mathtools}
\usepackage{float}
\usepackage{diagbox}
\usepackage{placeins}

\numberwithin{equation}{section}

\newcommand{\re}{\text{Re}\,}
\newcommand{\im}{\text{Im}\,}

\newcommand{\beq}{\begin{equation}}
\newcommand{\eeq}{\end{equation}}

\title{\boldmath Minimal seesaw and leptogenesis with the smallest modular finite group}
                                        	 \author{Simone Marciano,}
                                           \author{Davide Meloni,}
                                           \author{Matteo Parriciatu}

                                           \affiliation{Dipartimento di Matematica e Fisica, Università degli Studi Roma Tre, \\ INFN Sezione Roma Tre,\\
                                           Via della Vasca Navale 84,  00146 Rome, Italy}
                                      
                                       \emailAdd{simone.marciano@uniroma3.it}
                                          \emailAdd{davide.meloni@uniroma3.it}
                                           \emailAdd{matteo.parriciatu@uniroma3.it}
                                       
                                           \abstract{We propose a model for leptons based on the smallest modular finite group $\Gamma_2\cong S_3$ that, for the first time, accounts for both the hints of large low-energy CP-violation in the lepton sector and 
the matter-antimatter asymmetry of the Universe, generated by only two heavy right-handed neutrinos. These same states  are also employed in a Minimal seesaw mechanism to generate light neutrino masses. Besides the heavy neutrinos, our particle content is the same as the Standard Model (SM), with the addition of one single modulus $\tau$, whose vacuum expectation value is responsible for both the modular and CP-symmetry breakings.  We show that this minimalistic SM extension is enough to get an excellent fit to low energy neutrino observables and to the required baryon asymmetry $\eta_B$. Predictions for the neutrino mass ordering, effective masses in neutrinoless double beta decay and tritium decay as well as for the Majorana phases are also provided.}
                                           \arxivnumber{2402.18547}
                                           
\begin{document}
                                           \maketitle

\flushbottom

%

\setcounter{footnote}{0}
\setcounter{tocdepth}{1}

%

\renewcommand{\thefootnote}{\arabic{footnote}}
\setcounter{footnote}{0}
\setcounter{tocdepth}{1}

\section{Introduction}

The flavor structure of the Standard Model (SM) is still an outstanding puzzle in theoretical particle physics. The quite pronounced mass hierarchies between the three generations of charged fermions in both quark and lepton sectors are not addressed by the SM gauge symmetry, while the smallness of the neutrino mass scale calls for an entirely different mechanism beyond the SM. The mixing of the generations and CP-violation are also completely unconstrained by gauge symmetry. As we enter the precision era of neutrino experiments, we face the theoretical challenge to derive at least a major portion of the free parameters from an organizing principle. In particular, the peculiar mixing pattern of leptons (characterized by two large and one small angles), together with the recent hints of CP violation in neutrino oscillations could be understood with discrete symmetries through non-Abelian groups \cite{Altarelli:2010gt}. However, this approach proved to be quite unsatisfactory due to the unconstrained number of free parameters, and to the complicated scalar sector equipped with a substantial number of flavons (spurion fields) which are needed to reproduce the observed mixing \cite{Feruglio:2019ybq}. A step forward was made by Feruglio in 2017 \cite{Feruglio:2017spp} where modular symmetry takes the role of a flavor discrete symmetry in a bottom-up approach. The Yukawa couplings of the SM become modular forms: pre-determined functions of a complex field $\tau$ called “modulus", which describes the geometry of compactified extra-dimensions in superstring theory \cite{Lauer:1989ax,Lauer:1990tm}. As a result, models based on modular symmetry are in general more predictive as they require only a limited number of free parameters, and the flavon vacuum alignment problem is replaced by the moduli stabilization, at least in top-down approaches \cite{Ishiguro:2020tmo,Novichkov:2022wvg,Ishiguro:2022pde,Knapp-Perez:2023nty}. Both the usual matter superfields and Yukawa couplings transform in irreducible representations of the modular finite group $\Gamma_N$ of a given level $N\in\mathbb{N}$. The appealing feature is that for $N\le 5$ these groups are isomorphic to the more familiar non-Abelian discrete groups $S_3, A_4, S_4$ and $A_5$. As opposed to traditional flavor models based on those groups, in modular symmetry a generic vacuum expectation value (VEV) of the modulus breaks the modular group completely, and the observables of the lepton sector are completely determined by the value of this VEV, up to a limited number of free parameters in the superpotential. Currently, in the literature there is a substantial number of phenomenologically viable modular models for the lepton sector, some examples are given by: $\Gamma_2\cong S_3$ \cite{Kobayashi:2018vbk,Meloni:2023aru}, $\Gamma_3\cong A_4$ \cite{Feruglio:2017spp,Kobayashi:2018vbk,Kobayashi:2018wkl,Criado:2018thu,Kobayashi:2018scp,Okada:2018yrn,Okada:2019uoy,Ding:2019zxk,Kobayashi:2019xvz,Asaka:2019vev,Ding:2019gof,Zhang:2019ngf,King:2020qaj,Ding:2020yen,Asaka:2020tmo,Okada:2020brs,Yao:2020qyy,Okada:2021qdf,Nomura:2021yjb,Chen:2021prl,Nomura:2022boj,Gunji:2022xig,Devi:2023vpe}, $\Gamma_4\cong S_4$ \cite{Ding:2019gof,Penedo:2018nmg,Novichkov:2018ovf,deMedeirosVarzielas:2019cyj,Kobayashi:2019mna,King:2019vhv,Criado:2019tzk,Wang:2019ovr,Wang:2020dbp,Qu:2021jdy}, $\Gamma_5\cong A_5$ \cite{Criado:2019tzk,Novichkov:2018nkm,Ding:2019xna}. Models based on the double-cover $\Gamma_N'$ of the modular finite groups have also been put forward, some examples are \cite{Liu:2019khw,Liu:2020akv,Wang:2020lxk,Yao:2020zml,Novichkov:2020eep,Novichkov:2021evw,deMedeirosVarzielas:2022fbw,Ding:2022nzn}. For a more exhaustive list of references, see the recent reviews \cite{Kobayashi:2023zzc,Ding:2023htn}.
In addition to the flavor puzzle, the source of the asymmetry between matter and antimatter in the early universe remains an open issue. In this regard, once a seesaw mechanism is introduced to give mass to neutrinos, leptogenesis \cite{Fukugita:1986hr,Barbieri:1999ma,Davidson_2008,Buchmuller:2004nz,Pilaftsis:2003gt,Buchmuller:2005eh} becomes almost inevitable. Indeed, Sakharov's conditions \cite{Sakharov:1967dj} can be naturally satisfied in this context; in fact,  the seesaw mechanism predicts a violation of lepton number through processes involving Majorana neutrinos, as well as complex Yukawa couplings that induce CP violation. In addition, for a wide range of parameter values, these processes can occur out of thermal equilibrium.

In the supersymmetric modular symmetry business, a bunch of papers have already discussed the issues related to leptogenesis; in particular, models based on $\Gamma_3\cong A_4$ have been presented in \cite{Gogoi:2023jzl, Ding:2022bzs,Kang:2022psa,Mishra:2022egy,Dasgupta:2021ggp,Okada:2021qdf,Kashav:2021zir,Behera:2020sfe,Asaka:2019vev}, those on  $\Gamma_4\cong S_4$ in \cite{Qu:2021jdy,Wang:2019ovr} while the double covering group $\Gamma_5'\cong A_5'$ has been discussed  in \cite{Behera:2022wco}.

 In this paper, we contribute to the subject presenting a modular model based on $S_3$ capable of explaining the low-energy lepton observables and CP-violation as well as to account for the correct amount of the baryon asymmetry of the Universe (BAU), $\eta_B$. Our construction is
“minimal" in the sense that only two right-handed neutrinos (RN) are added to the particle content of the SM and that the unique modulus $\tau$ generates both the low energy CP-violating phase and the high energy CP violation needed for the leptogenesis mechanism to account for the BAU, without the addition of any other complex phases in the Yukawa matrices. Our results for $\eta_B$ are obtained solving the appropriate Boltzmann equations in the  $N_1-$dominated scenario, that we discuss in detail. A number of interesting predictions can be drawn from our model, and are related to the effective masses in the neutrinoless double beta decay and tritium decay (all below the current experimental values), as well as to the values of the Majorana phases. 
The remaining sections of this  paper are organized as follows: in section \ref{modfla} we recall the main ingredients of the modular symmetry and its use in the supersymmetric (SUSY) context; section \ref{mod} is devoted to the analytic description of our model, which is subsequently analyzed numerically in section \ref{num}; the relevant part of the present manuscript related to leptogenesis is carefully described in section \ref{lepto}; in \ref{concl} we draw our conclusions. Three appendices are supplied: in appendix \ref{sec:N1dominato} we discuss the solution of the Boltzmann equations in presence of two sterile neutrinos and justify the adopted $N_1-$dominated scenario; in appendix \ref{apforms} we report our choice for the {arbitrary} normalization of the level $N=2$ modular forms; finally, in the last appendix \ref{rescD} we show how to map the points outside the fundamental domain into points belonging to it.


\section{Modular flavor symmetry at level $2$}
\label{modfla}
The modular group $\overline\Gamma=\text{SL}(2,\mathbb{Z})/\{\pm\mathbbm{1}\}$ acts on the modulus $\tau$, restricted to the upper-half complex plane, through the transformation $\gamma\,:\, \tau\to \gamma(\tau)$,
\begin{equation}
\label{gammatau}
\gamma(\tau)=\frac{a\tau+b}{c\tau+d}\quad,\quad a,b,c,d\in\mathbb{Z}\quad, \quad ad-bc=1\,,
\end{equation}
which is generated by $S$ and $T$ defined as:
\begin{equation}
\label{essti}
\begin{cases}
S\,:\,\tau \rightarrow -\displaystyle\frac{1}{\tau} \\ \\
T\,:\,\tau\rightarrow \tau+1
\end{cases}\quad\quad,\quad\quad\quad S^2=(ST)^3=\mathbbm{1}\,.
\end{equation}
The generators \eqref{essti} are represented in $\text{SL}(2,\mathbb{Z})$ by the $2\times2$ matrices $
S=\begin{pmatrix}
0 &1\\
-1&0
\end{pmatrix}$, and $T=\begin{pmatrix}
1 &1\\
0&1
\end{pmatrix}$. A modular form is a holomorphic function of $\tau$ which, under $\overline{\Gamma}$, transforms as:
\begin{equation}
    f(\gamma(\tau))=(c\tau+d)^kf(\tau)\,,
\end{equation}
where $k$ is a positive integer called “{\it weight}".\newline
While the group $\overline\Gamma$ is infinite and non-compact, compact and finite groups can be constructed from its infinite normal subgroups $\Gamma(N)$ for $N=1,2,3...$ defined as:
\begin{equation}
\label{princN}
\Gamma(N)=\left\{\begin{pmatrix}
a& b\\
c&d
\end{pmatrix}\in \text{SL}(2,\mathbb{Z})\,\Big| \begin{pmatrix}
a& b\\
c&d
\end{pmatrix}\equiv\begin{pmatrix}
1& 0\\
0&1
\end{pmatrix}(\text{mod } N) \, \right\}\,,
\end{equation}
where the natural number $N$ is called the \textit{level}. In \cite{Feruglio:2017spp} Feruglio has shown that it is always possible to find a basis where modular forms of a given level transform in unitary representations of the finite groups $\Gamma_N\equiv \overline{\Gamma}/\overline{\Gamma}(N)$ which, for $N\le 5$, are isomorphic to the non-Abelian discrete groups $S_3,A_4,S_4,A_5$. Here we focus on the lowest level $N=2$, which labels the smallest finite modular group $\Gamma_2\cong S_3$. As reviewed in \cite{Meloni:2023aru}, modular forms of level $2$ and (even) weight $k$ span a linear space of finite dimension $k/2+1$, and a basis of lowest weight forms for all modular forms of level $2$ is provided by the $S_3$ doublet $Y_1(\tau),Y_2(\tau)$ given in appendix \ref{apforms}. Those can be employed to construct modular forms of higher weights using the $S_3$ composition rules.

In what follows, we will use the convenient notation:\footnote{In $Y_{\boldsymbol\rho}^{(a)}(\tau)$, the symbol $\boldsymbol{\rho}$ denotes the irrep, the upper index $(a)$ signals how many times the basis $Y_1(\tau),Y_2(\tau)$ was contracted to form $Y_{\boldsymbol\rho}^{(a)}(\tau)$, thus having weight $2a$. Individual components of the multiplets are simply given by $Y_{j}^p(\tau)$, where $j=\{1,2\}$ and $p$ is an exponent.}  
\begin{equation}
\label{ydoublets}
Y^{(1)}_\mathbf{2}(\tau)\equiv\begin{pmatrix}Y_1(\tau)\\
Y_2(\tau)
\end{pmatrix}_\mathbf{2}\,.
\end{equation}
Making use of the composition rules listed in appendix \ref{apforms}, the $3$ linearly independent modular forms of weight $4$ are constructed in the following way:
\begin{equation}
\label{weight2}
Y^{(1)}_\mathbf{2}(\tau)\otimes 
Y^{(1)}_\mathbf{2}(\tau)=(Y_1^2(\tau)+Y_2^2(\tau))_\mathbf{1}\oplus \begin{pmatrix}
Y_2^2(\tau)-Y_1^2(\tau)\\
2Y_1(\tau)Y_2(\tau)
\end{pmatrix}_\mathbf{2}\,,
\end{equation}
which, according to our notation, will be expressed as $Y^{(2)}_\mathbf{1}(\tau)$ and $Y^{(2)}_\mathbf{2}(\tau)$, respectively. Similarly, there are $4$ linearly independent modular forms of weight $6$:
\begin{equation}
\begin{split}
\label{sixweight}
Y^{(2)}_\mathbf{2}(\tau)\otimes Y^{(1)}_\mathbf{2}(\tau)=[3Y_1(\tau)Y_2^2(\tau)-Y_1^3(\tau)]_\mathbf{1}\oplus [Y_2^3(\tau)-3Y_2(\tau)Y_1^2(\tau)]_\mathbf{1'}\oplus\\ \oplus \begin{pmatrix}
Y_1(\tau)(Y_1^2(\tau)+Y_2^2(\tau))\\
Y_2(\tau)(Y_1^2(\tau)+Y_2^2(\tau))
\end{pmatrix}_\mathbf{2}\,,
\end{split}
\end{equation}
labelled in our notation as $Y^{(3)}_\mathbf{1}$, $ Y^{(3)}_\mathbf{1'}$ and $Y^{(3)}_\mathbf{2}$, respectively. Since the number of independent forms of weight $k$ grows with $k/2+1$, we stop at weight $6$ to aim at a minimalistic number of free parameters in the superpotential.
One of the appealing features of modular invariance is that the value of the modulus can be restricted to a \textit{fundamental} domain defined as:
\begin{equation}
\label{fun_domain}
\mathcal{D}=\left\{\tau\in \mathbb{C}\,:\, \text{Im}\,\tau>0\,,\, \,\,\left|\text{Re}\,\tau\right|\le \frac{1}{2}\,,\,\,\,|\tau|\ge 1 \right\}\,.
\end{equation}

This corresponds to the gray region $\mathcal{D}$ in figure \ref{Dmaps},  appendix \ref{rescD}. Values outside $\mathcal{D}$ are redundant and can be mapped inside \eqref{fun_domain} through the appropriate modular transformation \eqref{gammatau}. It is interesting to observe that, in particular for $\Gamma_2\cong S_3$, the modular forms of lowest weight $Y_1(\tau),Y_2(\tau)$ are intrinsically hierarchical for $\tau\in\mathcal{D}$, see eq.\eqref{qexp}; this allows us to define the appropriate ratio:
\begin{equation}
\label{epss}
    \frac{Y_2(\tau)}{Y_1(\tau)}\equiv \zeta =|\zeta|\,e^{ i\, g}\,,
\end{equation}
where $g=g(\re\tau)$ is a real function depending on $\re\tau$. The absolute value $|\zeta|$ satisfies $|\zeta|\lesssim 1$ for $\tau\in\mathcal{D}$ and is suppressed by $e^{-\displaystyle\pi \im\tau}$, as shown in figure \ref{epsmap}. Thus, we can use $Y_2(\tau)/Y_1(\tau)$ as an expansion parameter, independently of the chosen normalization of the modular forms (more about that in appendix \ref{rescD}).


\subsection{The SUSY framework}
The supersymmetric modular-invariant action, turning off gauge interactions for simplicity, is written as:
\begin{equation}
\label{action}
\mathcal{S}=\int d^4x\int d^2\theta d^2\bar\theta \,K(\Phi,\bar\Phi)+\left[\int d^4x\int d^2\theta\, \mathcal{W}(\Phi)+\text{h.c.}\right]\,,
\end{equation}
where $\theta,\bar\theta$ are the superspace coordinates; $\Phi=(\tau,\varphi)$ are chiral superfields, and $\varphi$ labels the usual matter supermultiplets; $K(\Phi,\bar\Phi)$ is the Kähler potential and $\mathcal{W}(\Phi)$ is the superpotential. We assume that chiral superfields transform under $\overline\Gamma$ as:
\begin{equation}
\label{moddd_transf}
\tau\to \gamma(\tau)=\displaystyle\frac{a\tau+b}{c\tau+d} \quad\quad,\quad\quad 
\varphi^{(I)}\to (c\tau+d)^{-k_I}\rho^{(I)}(\gamma)\varphi^{(I)}\,,
\end{equation}
where $\gamma\in\overline{\Gamma}$ and $\rho^{(I)}(\gamma)$ is the matrix for the irreducible representation of $\Gamma_N$. Here the $k_I$ of the superfields are called modular charges. The action \eqref{action} is invariant under \eqref{moddd_transf} provided that the superpotential $\mathcal{W}(\Phi)$ and the Kähler potential transform as:
\begin{equation}
\label{ccaler}
\begin{cases}
\mathcal{W}(\Phi)\to\mathcal{W}(\Phi)\\ \\
K(\Phi,\bar\Phi)\to K(\Phi,\bar\Phi)+f(\Phi)+\bar{f}(\bar\Phi)
\end{cases}
\end{equation}
with $f(\Phi)$ being holomorphic. The superpotential $\mathcal{W}(\Phi)$ will generally be a combination of Yukawa couplings $Y_{I_1...I_n}(\tau)$ and chiral superfields
\begin{equation}
\label{gen_super}
\mathcal{W}(\Phi)=\sum (Y_{I_1...I_n}(\tau)\,\varphi^{(I_1)}...\varphi^{(I_n)})_\mathbf{1}\,.
\end{equation}
The Yukawa couplings are assumed to be modular forms which transform as:
\begin{equation}
\label{yuk_transf}
Y_{I_1...I_n}(\gamma(\tau))=(c\tau+d)^{k_Y}\rho(\gamma)Y_{I_1...I_n}(\tau)\,,
\end{equation}
with $k_Y\ge 2$ being an even integer in the case of $N=2$. The novelty compared to classic non-Abelian discrete groups model building is that the invariance of \eqref{gen_super} is achieved by satisfying not only the existence of a singlet contraction $\rho\otimes \rho_{I_1}\otimes \rho_{I_2}...\otimes \rho_{I_n} \supset\mathbf{1}$ between all the irreps involved, but also that every operator in \eqref{gen_super} must be \textit{weightless}, i.e. $k_Y=k_{I_1}+k_{I_2}+...+k_{I_n}$, the total weight of the matter fields must be counterbalanced by appropriate Yukawa modular forms of weight $k_Y$.\newline
Finally, in most economical realizations of modular flavor models, the invariant Kähler potential is given by:
\begin{equation}
\label{kahler}
K(\Phi,\bar\Phi)=-h\Lambda_\tau^2 \log(-i\tau+i\bar\tau)+\sum_I(-i\tau+i\bar\tau)^{-k_I}{|\varphi^{(I)}|^2}\,,
\end{equation}
where $h>0$ and $\Lambda_\tau$ has mass-dimensions of one. Note that we used the same symbol to denote the superfield and its scalar component. 

In what follows, the vacuum expectation value (VEV) of the modulus $\tau$ is the only source of flavor symmetry breaking, thus no additional flavons are needed.\footnote{Here we do not deal with the dynamical justification for this VEV, which is a subject of ongoing studies. See for examples \cite{Kobayashi:2019xvz}, and \cite{Novichkov:2022wvg,Knapp-Perez:2023nty,Ishiguro:2020tmo,Ishiguro:2022pde}.} The scale of modular symmetry breaking can be located at the compactification scale, which is supposed to be near the Planck scale. The modular group is completely broken for every value of $\tau$ except for the so-called \textit{symmetric} points $\tau_\text{sym}=\{\omega,i,i\infty\}$ where $\omega\equiv e^{2\pi i/3}$, which are left invariant under $ST$, $S$, and $T$ respectively \cite{Novichkov:2018ovf}.

\subsection{gCP symmetry \label{ggcpp}}
In order to reduce the number of model free parameters (thus possibly making the model itself more predictive), we assume a generalized CP symmetry (gCP). This was made consistent with modular symmetry in \cite{Novichkov:2019sqv} where it was shown that under certain requirements for the group basis and the normalization of the Yukawa forms, imposing a gCP symmetry consists in the reality condition of the superpotential couplings.

Given this assumption, the only source of CP violation in gCP models is the VEV of the complex modulus. Since $\tau$ appears in the expansions \eqref{qexp} of appendix \ref{apforms} through powers of $e^{2\pi i\tau}$, the CP violation is controlled entirely by $\re\tau$ for $\tau\in\mathcal{D}$. This means that the only complex phase in the mass matrices is represented by powers of $e^{2\pi i\,\re\tau}$. Apart from the obvious $\re\tau=0$, the other CP-conserving points consist in the boundary of $\mathcal{D}$. Note that all the previously mentioned symmetric points are CP-conserving.


\section{The model}
\label{mod}
Given that our main concern here is to find a viable seesaw model for generating neutrino masses, for the charged-leptons sector we simply employ the transformation properties under $\Gamma_2$ and a charge assignment compatible with {\it naturally} hierarchical charged leptons, adopted from the construction illustrated in \cite{Meloni:2023aru}.
In this respect, the superfields $E_i^c$ correspond to the three flavors of right-handed charged leptons (respectively $\{i=1,2,3\}\equiv \{e,\mu,\tau\}$) with modular charges $4,0$ and $-2$, respectively. 
The first two lightest left-handed $SU(2)_\text{L}$ doublets
are grouped into a doublet of $S_3$ while the heaviest family belongs to a pseudo-singlet.
Finally, the Higgs doublets transform as invariant singlets with modular charges set to $0$.  
In table \ref{irrepss} we summarize 
the chiral supermultiplets of our model, the transformation properties under $\Gamma_2\cong S_3$ and the related modular charges.
As a consequence, the allowed terms in the superpotential are the following:
\begin{equation}
\label{ncl}
\mathcal{W}_e^{H}=\alpha E_1^cH_d(D_\ell Y_\mathbf{2}^{(3)})_\mathbf{1}+\beta E_2^cH_d(D_\ell Y_\mathbf{2})_\mathbf{1'}+\gamma E_3^cH_d\ell_3+\alpha_D E_1^cH_d\ell_3 Y_\mathbf{1'}^{(3)}\,,
\end{equation}
and the resulting mass matrix (in the right-left basis) after electroweak symmetry breaking reads\footnote{In the parentheses notation $(...)_{1,2}$ we denote the two components of the corresponding doublet $Y_{\mathbf{2}}^{(a)}$.}:
\begin{equation}
\label{mass_ncl}
M_\ell=\begin{pmatrix}
\alpha (Y^{(3)}_\mathbf{2})_1 & \alpha(Y^{(3)}_\mathbf{2})_2 & \alpha_D Y^{(3)}_\mathbf{1'} \\ \beta Y_2&-\beta Y_1& 0 \\ 0 & 0 & \gamma
\end{pmatrix}_\text{RL}v_d\,,
\end{equation}
where $v_d$ is the VEV of $H_d$ and the modular forms of weight $6$ are given by \eqref{sixweight}. One can recast the matrix \eqref{mass_ncl} in a form suitable for getting perturbative expressions of the eigenvalues; in order to do that, we define 
$A\equiv \alpha_D/\alpha$, $B\equiv (\beta/\alpha)Y_1^{-2}$, $C\equiv (\gamma/\alpha)Y_1^{-3}$ and $\zeta$ is the expansion parameter defined in \eqref{epss}. Thus, we get:
\begin{equation}
\label{mass_par}
M_\ell=v_d\alpha Y_1^3\begin{pmatrix}
(1+\zeta^2) & \zeta(1+\zeta^2) & -A\zeta(3-\zeta^2) \\ B \zeta&-B& 0 \\ 0 & 0 & C
\end{pmatrix}_\text{RL}\,.
\end{equation}
Assuming $\frac{\beta}{\alpha}\sim\frac{\gamma}{\alpha}\approx 1$ and using the $q$-expansions \eqref{qexp} we estimate $|B|,|C|\gg A$ for $\tau\in\mathcal{D}$. It is now easy to obtain the approximated eigenvalues in powers of $\zeta$ as follows:
\begin{align}
\label{eigenhierarc}
&m_e=v_d\alpha \left(|Y_1^3|+\frac{3}{2}|Y_1^3||\zeta|^2+\mathcal{O}(\zeta^3)\right)\\
&m_\mu=v_d\alpha \left(|Y_1|+\frac{1}{2}|Y_1| |\zeta|^2+\mathcal{O}(\zeta^3)\right)\\
&m_\tau=v_d\alpha \left(1+\frac{9A^2}{2}|Y_1^6||\zeta|^2+\mathcal{O}(\zeta^3)\right)\,.
\end{align}
The hierarchy $(m_\tau,m_\mu,m_e)\sim m_\tau(1,|Y_1|,|Y_1|^3)$ naturally arises considering that  $|Y_1|\approx 7/100$. 
Given the non-diagonal structure of the charged-leptons mass matrix \eqref{mass_par}, a non-trivial contribution to the PMNS matrix arises; considering that $A/|C|^2\ll |\zeta|$, the unitary matrix $U_\ell$ that diagonalises $M_\ell^\dagger M_\ell$ can be expressed as follows:
\begin{equation}
\label{unitarymin}
U_\ell\sim\begin{pmatrix}
1-\displaystyle\frac{|\zeta|^2}{2}& -\overline{\zeta} &0\\
\zeta & 1-\displaystyle\frac{|\zeta|^2}{2} &0\\
 0& 0& 1
\end{pmatrix}+\mathcal{O}(\zeta^3)\,,
\end{equation}
up to $\mathcal{O}(1)$ coefficients. In particular, we see that $U_\ell$ would provide a non-negligible correction to the solar angle, for which we expect only a small $\mathcal{O}(\zeta)$ contribution for $\text{Im}\,\tau$ sufficiently large. 
All in all, the hierarchical texture of the charged-leptons is provided by the modular invariance through our construction. The mass spectrum is completely determined by five dimensionless real parameters, $\text{Re}\,\tau$, $\text{Im}\,\tau$, $\alpha_D/\alpha$, $\beta/\alpha$ and $\gamma/\alpha$ and one global scale $v_d\alpha$.


\subsection{Neutrino sector}
\begin{table}[t]
\aboverulesep = 0pt
\belowrulesep = 0pt
\centering
\def\arraystretch{1.3}
\resizebox{0.9\textwidth}{!}{
\begin{tabular}{@{}|l|c|c|c|c|c|c|c|@{}}
\toprule
\multicolumn{1}{|c|}{} &
  \multicolumn{1}{c|}{$E_1^c$} &
  \multicolumn{1}{c|}{$E_2^c$} &
  \multicolumn{1}{c|}{$E_3^c$} &
  \multicolumn{1}{c|}{$D_\ell$} &
  \multicolumn{1}{c|}{$\ell_3$} &
  \multicolumn{1}{c|}{$H_{d,u}$} &
   \multicolumn{1}{c|}{$N^c$} 
\\ \midrule
$\scriptstyle{SU(2)_L\times U(1)_Y}$ &  $(\mathbf{1},+1)$       & $(\mathbf{1},+1)$       & $(\mathbf{1},+1)$        & $(\mathbf{2},-1/2)$    & $(\mathbf{2},-1/2)$ &   $(\mathbf{2},\mp1/2)$ & $(\mathbf{1},0)$ \\ \midrule
$\Gamma_2\cong S_3$ &  $\mathbf{1}$     & $\mathbf{1'}$       & $\mathbf{1'}$       & $\mathbf{2}$        & $\mathbf{1'}$        & $\mathbf{1}$  & $\mathbf{2}$  \\ \midrule 
$k_I$ &  $4$       & $0$       & $-2$       & $2$        & $2$        & $0$ & $2$  \\  \bottomrule 
\end{tabular}}
\caption{\small{\it Chiral supermultiplets, transformation properties under $\Gamma_2\cong S_3$ and modular charges. For the sign convention of the modular charges, we refer to our choice in eq. \eqref{moddd_transf}.}}
\label{irrepss}
\end{table}
To keep the number of free parameters at minimum, we adopt the Minimal seesaw scenario \cite{King:1999mb,Frampton:2002qc,King:2002qh,Raidal:2002xf,King:2002nf} consisting in the introduction of two SM singlets right-handed Majorana fields $N^c_1,N^c_2$. As listed in table \ref{irrepss}, these are assigned to an $S_3$ doublet $N^c\sim\mathbf{2}$. The invariant superpotential of the neutrino sector is given by:
\begin{equation}
\label{neutrino_sup}
\begin{split}\mathcal{W}_\nu=gH_uN^cD_\ell Y^{(2)}_\mathbf{2}+g'H_u(N^cY^{(2)}_\mathbf{2})_\mathbf{1'}\ell_3+g''H_u(N^cD_\ell)_\mathbf{1}Y^{(2)}_\mathbf{1}+\\ 
+\Lambda [(N^cN^c)_\mathbf{2}Y^{(2)}_\mathbf{2}+\lambda(N^cN^c)_\mathbf{1}Y^{(2)}_\mathbf{1}]\,,
\end{split}
\end{equation}
where $g,g',g'',\lambda$ are dimensionless free parameters and $\Lambda$ is the Majorana right-handed mass scale. As mentioned in section \ref{ggcpp}, all these couplings (as well as the ones from the charged-leptons sector) are real due to the imposed gCP symmetry. The only complex parameter is the modulus $\tau$.

Exploiting the composition rules of $S_3$ given in appendix \ref{apforms}, the Dirac mass matrix reads:
\begin{equation}
\label{mdirac}
M_D= gv_u\begin{pmatrix}
-(Y_2^2-Y_1^2)+\displaystyle\frac{g''}{g}(Y_1^2+Y_2^2)&2Y_1Y_2&\displaystyle\frac{g'}{g}(2Y_1Y_2)\\ 
2Y_1Y_2 & (Y_2^2-Y_1^2)+\displaystyle\frac{g''}{g}(Y_1^2+Y_2^2) &- \displaystyle\frac{g'}{g}(Y_2^2-Y_1^2)
\end{pmatrix}_\text{RL}\,,
\end{equation}
where $v_u$ denotes the VEV of the neutral scalar component of $H_u$. The Majorana mass matrix of right-handed neutrinos is given by:
\begin{equation}
\label{mmajo}
\mathcal{M}_R=\Lambda\begin{pmatrix}
-(Y_2^2-Y_1^2)+\lambda (Y_1^2+Y_2^2)&2Y_1Y_2\\
2Y_1Y_2&(Y_2^2-Y_1^2)+\lambda (Y_1^2+Y_2^2)
\end{pmatrix}_\text{RR}\,.
\end{equation}
The mass matrix of the light Majorana neutrinos is then obtained from the well known formula of type-I seesaw:
\begin{equation}
\label{numass}
    m_\nu=-M_D^T\mathcal{M}_R^{-1}M_D\,.
\end{equation}
Note that, as a consequence of the Minimal seesaw, the matrix $m_\nu$ has rank $2$, i.e. in the limit of exact SUSY our model predicts that the lightest neutrino is massless.\footnote{Regarding the stability of $m_\textit{lightest}=0$ against the quantum corrections, it has been shown that the condition still holds at one-loop for a running between the seesaw and electroweak scales. The two-loop effect is vanishingly small \cite{Xing:2020ald}.} From \eqref{numass} it follows that the mass-scale of light neutrinos is completely determined by the global parameter $g^2v_u^2/\Lambda$ and, from the discussion below eq. \eqref{unitarymin}, we conclude that the low-energy neutrino mixing and mass-splittings will be mainly dictated by five parameters: $\re\tau,\im\tau, g'/g, g''/g, \lambda$. Note that, among these, the real and imaginary part of $\tau$ already appear in the charged-leptons mass matrix.

\section{Numerical analysis and results \label{numres}}
\label{num}
\begin{table}[t]
\centering
\renewcommand{\arraystretch}{1.2}
\begin{tabular}{l c c} 
\toprule
Parameter$\qquad\qquad$ & \multicolumn{2}{c}{Best-fit value and $1\sigma$ range} \\ 
\midrule
& \textbf{NO} & \textbf{IO} \\

$r \equiv \Delta m^2_\text{sol}/|\Delta m^2_\text{atm}|$ & $0.0295\pm0.0008$ & $0.0298\pm0.0008$\\
$\sin^2\theta_{12}$ & $0.303^{+0.012}_{-0.012}$ & $0.303^{+0.012}_{-0.011}$ \\
$\sin^2\theta_{13}$ & $0.02225^{+0.00056}_{-0.00059}$ & $0.0223^{+0.00058}_{-0.00058}$ \\
$\sin^2\theta_{23}$ & $0.451^{+0.019}_{-0.016}$ & $0.569^{+0.016}_{-0.021}$ \\
$J_{\text{CP}}$ & $-0.027_{-0.010}^{+0.010}$ & $-0.032_{-0.007}^{+0.007}$ \\
\midrule
$m_e / m_\mu$ & \multicolumn{2}{c}{$0.0048 \pm 0.0002$} \\
$m_\mu / m_\tau$ & \multicolumn{2}{c}{$0.0565 \pm 0.0045$} \\ 
\bottomrule
\end{tabular}
\caption{\small{\it{Neutrino observables and their 1$\sigma$ ranges, from NuFIT 5.2 \cite{Esteban:2020cvm} using the dataset with SK atmospheric data \cite{Super-Kamiokande:2017yvm}. Here “Normal ordering" and “Inverted ordering" for the mass spectrum are indicated with \textbf{NO} and \textbf{IO}, respectively. The mass ratios of charged-leptons are taken from \cite{Feruglio:2017spp}. Here we defined $\Delta m^2_\text{sol}\equiv m_2^2-m_1^2$, $|\Delta m^2_\text{atm}|\equiv |m_3^2-(m_1^2+m_2^2)/2|$. The Jarlskog invariant is defined in \eqref{jarls} and can be expressed as $J_\text{CP}=c_{12}s_{12}c_{23}  s_{23} c_{13}^2s_{13}\sin\delta_\text{CP}$. To extract its best-fit value and $1\sigma$ uncertainty, we referred to its one-dimensional $\chi^2$ projection from NuFIT 5.2. The charged-leptons mass ratios are taken from \cite{Feruglio:2017spp}. The running of low-energy neutrino observables can be neglected since in our fit we do not obtain a quasi-degenerate spectrum \cite{Antusch:2003kp}.}
}}
\label{nufit_table}
\end{table}
In total, the parameter space of our model consists of only eight dimensionless parameters, as discussed above.\footnote{The overall mass scales $\alpha v_d$ and $g^2v_u^2/\Lambda$ are easily recovered through the values of $m_\tau$ and $|\Delta m_\text{atm}^2|$ and thus are not introduced into the fit. In doing so, we used for $m_\tau$ the value reported in \cite{Yao:2020qyy} and $|\Delta m^2_\text{atm}|/(10^{-3}\text{ eV}^2)=2.507^{+0.026}_{-0.027}\,\textbf{(NO)}, \,\, \,\, 2.486^{+0.025}_{-0.028}\,\textbf{(IO)}$ from NuFIT 5.2.} To verify that our model is able to reproduce the seven experimental dimensionless observables reported in table \ref{nufit_table}, we performed a $\chi^2$ analysis for which we used the Gaussian approximation
\begin{equation}
\label{chisq}
\chi^2(p_i)=\sum_{j=1}^7 \left(\frac{q_j(p_i)-q_j^{\text{b-f}}}{\sigma_j}\right)^2\,,
\end{equation}
for the mixing $\{\sin^2\theta_{12},\sin^2\theta_{13},\sin^2\theta_{23}\}$, the mass ratios $\{m_e/m_\mu, m_\mu/m_\tau, r\}$ and for the invariant $J_\text{CP}$. Here $q_j(p_i)$ are the observables obtained from our model with the set of dimensionless parameters $p_i=\{\beta/\alpha,...,g'/g,...,\lambda\}$ taken as input. On the other hand, $q_j^{\text{b-f}}$ and $\sigma_j$ are the best-fit values and the $1\sigma$ symmetrized uncertainties taken from table \ref{nufit_table}.
The Jarlskog invariant $J_\text{CP}$ parameterises the amount of CP violation in the lepton sector and is defined as \cite{Esteban:2020cvm}:
\begin{equation}
\label{jarls}
    J_\text{CP}=\im[U_{11}U_{12}^*U_{21}^*U_{22}]\,,
\end{equation}
where $U\equiv U_\ell^\dagger U_\nu$ is the PMNS matrix \cite{Maki:1962mu,Pontecorvo:1957qd} for the lepton mixing, which in the case of Majorana neutrinos can be written in the standard convention:
\begin{equation}
\label{pmns_full}
U =
\begin{pmatrix}
c_{12} c_{13} & s_{12} c_{13} & s_{13}e^{-i\delta _\text{CP}} \\
-s_{12} c_{23} - c_{12} s_{23} s_{13}e^{i\delta _\text{CP}} &
c_{12} c_{23} - s_{12} s_{23} s_{13}e^{i\delta _\text{CP}} & s_{23} c_{13} \\
s_{12} s_{23} - c_{12} c_{23} s_{13}e^{i\delta _\text{CP}} &
-c_{12} s_{23} - s_{12} c_{23} s_{13}e^{i\delta _\text{CP}} & c_{23} c_{13}
\end{pmatrix}
\text{diag}(e^{-i\alpha_1},e^{-i\alpha_2},1)\,,
\end{equation}
where $c_{ij}\equiv \cos\theta_{ij}$ and $s_{ij}\equiv \sin\theta_{ij}$ and $\alpha_1,\alpha_2$ are the two Majorana phases. Note that, if the lightest neutrino is massless (as in our case), only one Majorana phase is physical.

\begin{figure}[H]
  \begin{center}
    \includegraphics[scale=0.21]{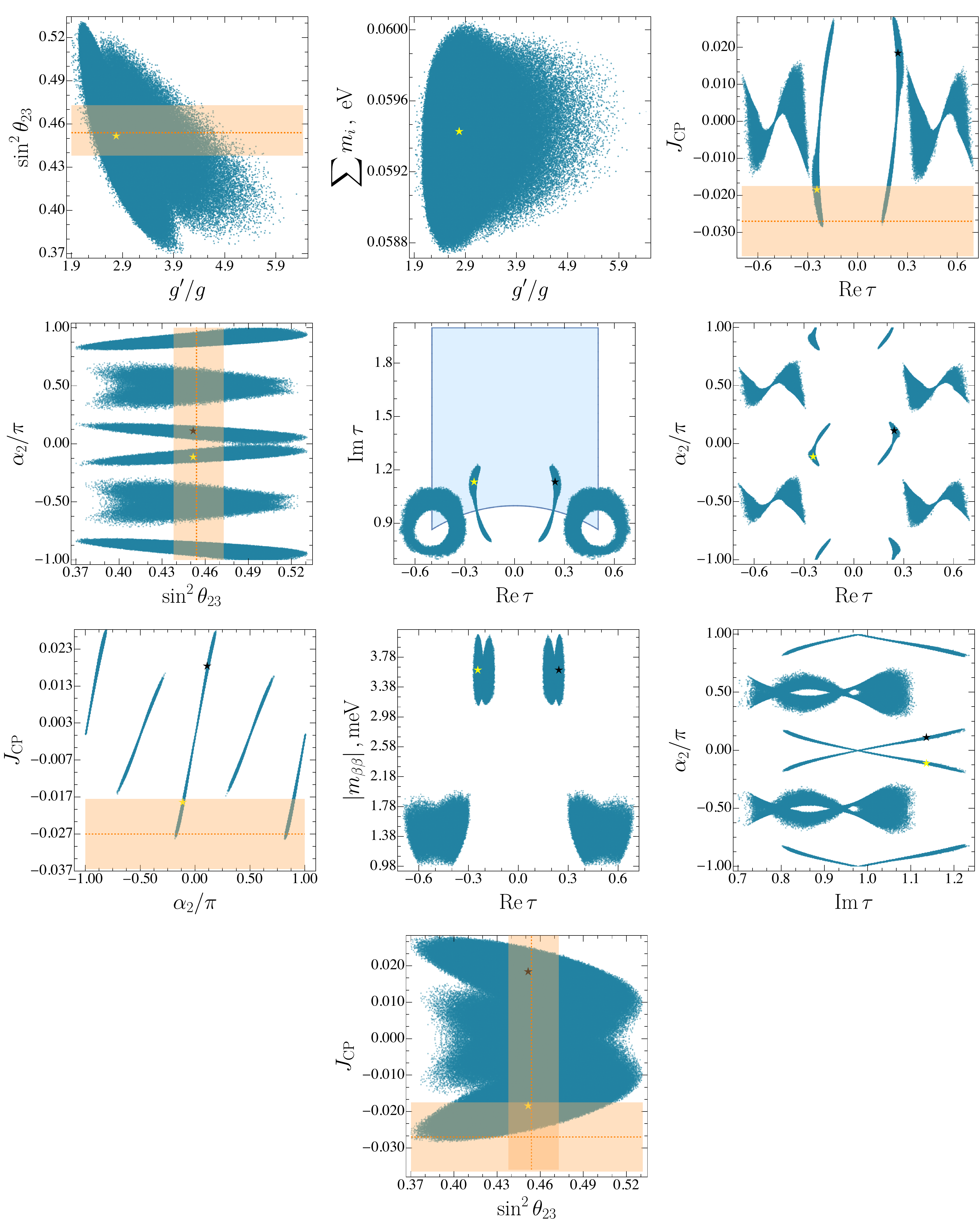}
    \caption{\small{\it Correlations between observables and free parameters. All the plotted points were accepted by the algorithm used for the fit and satisfy $\sqrt{\chi^2}\le 5$. The yellow point corresponds to the minimum $\chi^2$. The black point is the CP-transformed value of the modulus, which produces the inversion of all the CPV phases in the mass matrices. Orange lines and bands correspond to the experimental best fit and 1$\sigma$ ranges taken from table \ref{nufit_table}. Points located outside the fundamental domain are displayed for the purpose of illustration.}}
    \label{SS2plots}
  \end{center}
  \end{figure}

It must be noted that, given the CP symmetry of the model which consists in the invariance under $\tau\to -\tau^*$ (a reflection across the imaginary axis), we obtain two sets of points distinguished only by $\pm \re\tau$ with the CP violating (CPV) phases given by $\{\pm \delta_\text{CP}, \pm\alpha_1, \pm \alpha_2\}$. For thoroughness, we reported both sets in figure \ref{SS2plots}. The results of our fit\footnote{We explored the parameter space with an algorithm inspired by ref. \cite{Novichkov:2021cgl} which mimics the Brownian motion of a particle in a potential that depends on the figure of merit $l(p_i)=\sqrt{\chi^2(p_i)}$, where $p_i$ is the array of free dimensionless parameters mentioned in the text.} are given in table \ref{seesaw_table}, where the $1\sigma$ ranges correspond to the increase of one unit of the one dimensional $\chi^2$ projections from $\chi^2_\text{min}$.


\begin{table}[H]
\centering
\renewcommand{\arraystretch}{1.2}
\begin{tabular}{c|ccc}
   \toprule
  & Best-fit and $1\sigma$ range \\
  \midrule
  $\text{Re}\, \tau$ & $\pm 0.244_{-0.067}^{+0.012}$  \\
  $\text{Im}\,\tau$ & $1.132_{-0.297}^{+0.027}$  \\
  $\beta/\alpha$ & $0.92_{-0.03}^{+0.85}$  \\
  $\gamma/\alpha$ & $-1.20_{-2.14}^{+0.06}$  \\
  $\log_{10}(\alpha_D/\alpha)$ & $-13.4_{-76.3}^{+13.2}$  \\
  $g'/g$ & $2.76_{-0.23}^{+0.21}$  \\
  $g''/g$ & $-2.53_{-0.03}^{+0.13}$ \\
  $\log_{10}(|\lambda|)$ & $-12.2_{-59.2}^{+10.9}$ \\
  $v_d\,\alpha$, [GeV] & $1.08_{-0.69}^{+0.06}$ \\
  $v_u^2\, g^2 / \Lambda$ [eV] & $3.46_{-1.65}^{+0.55}$ \\
  \midrule
  $\sin^2 \theta_{12}$ & $0.305_{-0.011}^{+0.011}$  \\
  $\sin^2 \theta_{13}$ & $0.0221_{-0.0005}^{+0.0006}$   \\
  $\sin^2 \theta_{23}$ & $0.448_{-0.016}^{+0.014}$ \\
  $r$ & $0.0296_{-0.0008}^{+0.0006}$ \\
  $m_e/m_{\mu} $ & $0.0048_{-0.0002}^{+0.0001}$ \\
  $m_{\mu} / m_{\tau}$ & $0.0574_{-0.0050}^{+0.0032}$  \\
  \midrule
  Ordering & \textbf{NO} \\
  $J_\text{CP}$ & $ -0.018_{-0.002}^{+0.002}$   \\
  $\alpha_{1}/\pi$ & $0$  \\
  $\alpha_{2}/\pi$ & $\pm 0.112_{-0.014}^{+0.792}$  \\
  $m_1$ [meV] & $0$ \\
  $m_2$ [meV] & $8.620_{-0.123}^{+0.095}$ \\
  $m_3$ [meV] & $50.806_{-0.021}^{+0.016}$  \\
  $\textstyle \sum_i m_i$ [eV] & $0.0594_{-0.0001}^{+0.0001}$  \\
  $|m_{\beta\beta}|$ [meV] & $3.61_{-0.09}^{+0.09}$   \\
   $m_{\beta}^{\text{eff}}$ [meV] & $8.90_{-0.09}^{+0.10}$  \\

    \midrule
$d_\text{FT}$ & 3.03 \\
  \midrule
  $\chi^2_\text{min}$ & 0.98 \\
  \bottomrule
\end{tabular}
\caption{
\small{\it Best fit values of the free parameters and observables with their $1\sigma$ ranges, obtained as described in the text.}}
\label{seesaw_table}
\end{table}

 The fit is excellent, with $\chi^2_\text{min}=0.98$ and most dimensionless free parameters being $\mathcal{O}(1)$ with only two exceptions, related to the parameters $\alpha_D/\alpha$ (which can take values arbitrarily close to zero with a spread reaching $\mathcal{O}(1)$ values, all inside its $1\sigma$ range) and $\lambda$, which takes values arbitrarily close to zero as selected by the fit procedure. The degree of “fine-tuning" for the free parameters of the model, has been evaluated by means of a normalised Altarelli-Blankenburg measure $d_\text{FT}$ \cite{Altarelli:2010at}: 
\begin{equation}
    \label{fine_tuning}
    d_\text{FT}=\frac{\displaystyle\sum_i\left|\frac{\text{par}_i}{\delta\text{par}_i}\right|}{\displaystyle\sum_i\left|\frac{\text{obs}_i}{\sigma_i}\right|}\,.
\end{equation}
Here $|\text{par}_i|$ is the best-fit value of the free parameter and $|\delta\text{par}_i|$ is the shift of the parameter from the best-fit value which causes $\chi^2-\chi^2_\text{min}=1$, while keeping all the other parameters fixed to their best-fit values. Intuitively, a model can be considered fine-tuned if it takes even a very small shift from the best-fit value of some parameters to produce a large variation in the $\chi^2$. The measure \eqref{fine_tuning} is normalized by the absolute sum of the ratios between the observables best-fit values and their uncertainties. For the fit shown in table \ref{seesaw_table} we obtained $d_\text{FT}\sim\mathcal{O}(1)$ which suggests that our model is not affected by a fine-tuning much larger than the one which already affects the input data.

Let us now turn our attention to the predictions of our model. First of all, we found that the Inverted Ordering of the neutrino mass eigenstates is strongly disfavoured compared to the Normal Ordering (\textbf{NO}); equally relevant is the fact that, as anticipated, the lightest neutrino mass is exactly vanishing. This allows us to easily understand the numerical results on the Majorana effective mass of neutrinoless double-beta decay and the effective neutrino mass of tritium decay. In the first case, the usual definition (inserting $\alpha_1=0$) gives: 
\begin{equation}
\label{mbetabeta}
|m_{\beta\beta}|=\big|\sum_i m_i U_{ei}^2\big|=\sqrt{\Delta m^2_\text{atm}}\big|\sqrt{r}\,s_{12}^2c_{13}^2+\sqrt{1+r}\,s_{13}^2e^{-i(\delta_\text{CP}-\alpha_2)}\big|\sim \mathcal{O}(1)\,\text{meV}\,,
\end{equation}
while in the latter case:
\begin{equation}
\label{mtritium}
m_\beta^{\text{eff}}=\sqrt{\sum_i m_i^2|U_{ei}|^2}=\sqrt{\Delta m^2_\text{atm}}\,\sqrt{s_{13}^2+r(1-c_{13}^2c_{12}^2)}\sim \mathcal{O}(10)\,\text{meV}\,,
\end{equation}
in perfect agreement with the results shown in table \ref{seesaw_table}. Finally, the non-vanishing neutrino mass eigenstates lie in the region of tens of meV, implying $\sum_i m_i \sim 0.06$ eV, perfectly compatible with the most recent upper bound of $0.115\,\text{eV}$ ($95\,\%\,\text{C.L.}$) from \cite{eBOSS:2020yzd}. This, again, is mainly dictated by the \textbf{NO} scenario with vanishing lightest neutrino mass:
\begin{equation}
    \label{cosmo_sum_no}
    \sum_{i=1}^3 m_i= \sqrt{\Delta m^2_\text{atm}}(\sqrt{r}+\sqrt{1+r})\sim 0.06\,\text{eV}\,.
\end{equation}
In figure \ref{SS2plots} some correlations between observables (both fitted and predicted ones) and free parameters are displayed. All points in the plots have been selected by our algorithm and satisfy the relation $\sqrt{\chi^2}\le 5$. The yellow point corresponds to the minimum $\chi^2$ reported in table \ref{seesaw_table} while the black one is the CP-transformed value of the modulus, which produces the inversion of all the CPV phases in the mass matrices. The orange lines and bands correspond to the experimental best fit and 1 $\sigma$ ranges, respectively, taken from table \ref{nufit_table}. For illustrative purposes, we also report points located outside the fundamental domain.





\section{Leptogenesis \label{lepto_section}
}

\label{lepto}
We now discuss the mechanism of leptogenesis. It provides an elegant explanation for the matter-antimatter asymmetry of the Universe, attributing it to a lepton number asymmetry generated by the decays of heavy right-handed neutrinos. This asymmetry is then converted into a baryon asymmetry through non-perturbative processes referred to as \emph{sphalerons} \cite{GIUDICE200489,Buchmuller:2004nz,tHooft:1976snw}.

As commonly done, we express the baryon asymmetry by the parameter $\eta_B$, as
\begin{equation}
  \eta_B\equiv \frac{n_B-n_{\bar{B}}}{n_\gamma}\bigg|_0\; ,  
\end{equation}
where $n_B,n_{\bar{B}}$ and $n_\gamma$ refer to the number densities of the baryons, antibaryons and photons, while the subscript “0" means “at present time".

We have worked so far in the limit of exact supersymmetry. 
We will assume that the SUSY breaking scale $m_{SUSY}$ is above the mass scale of the right-handed neutrinos of our model, so that we can avoid considering the superparticles contribution to the final baryon asymmetry, as well as all the complications related to the gravitino densities \cite{Davidson:2002qv,GIUDICE200489,Buchmuller:2005eh,Pradler:2006qh}.
Nevertheless, it is imperative to exercise caution regarding the magnitude of $m_\text{SUSY}$, as arbitrarily large values may introduce complications. This is discussed in depth in Ref. \cite{Criado:2018thu}. Notably, the lepton masses undergo a correction which is proportional to the ratio $m_\text{SUSY}/\mathcal{F}$, where $\mathcal{F}$ represents the characteristic scale at which the supersymmetry-breaking sector communicates with the visible sector. This scale can be chosen in proximity to the Planck scale $M_{pl}$.
As we will discuss in detail in the following, our model provides right-handed neutrinos with masses of order $\mathcal{M}\sim \mathcal{O}(10^{12})$ GeV. Thus, we could choose $m_\text{SUSY}\simeq 10^{14}$ GeV and $\mathcal{F}\simeq 10^{18}$ GeV so that the corrections to the low energy observables do not affect our results, and we can proceed in discussing the leptogenesis without considering any massive superparticle.\\
In this section, we will  focus on  the so-called $N_1$ dominated scenario (N1DS) \cite{Samanta_2020,Blanchet_2012,Blanchet_2006}, even though our model involves one additional sterile state $N_2$. Then, we will solve numerically the semi-classical Boltzmann Equations \cite{Buchmuller:2004nz,Fong:2012buy,Davidson_2008} for the baryon asymmetry evolution, and we will present a realization that provides the value of the baryon-to-photon ratio $\eta_B$ inside the $3\sigma$ experimental allowed region \cite{2020}.

\subsection{Thermal leptogenesis: basic physics} 
\begin{figure}[t]\centering
    \begin{tikzpicture}[baseline=(current bounding box.center)]
        \begin{feynman}
          \node [dot] (a);
            \node [left=1.8cmof a] (b);
            \node [right=1 cmof b] (bb); 
            \node [above = 0.35cmof bb] (name1) {\large{$N_i$}};
            \node [above right=1.7cmof a] (c) {\large{$H_u$}};
            \node [below right=1.7cmof a] (d) {\large{$l_\alpha$}};
            \diagram* {
                (b) -- [plain,thick] (a);
                (a) -- [scalar,thick] (c),
                (a) -- [fermion,thick] (d)
            };
            \end{feynman}
    \end{tikzpicture}
     \begin{tikzpicture}[baseline=(current bounding box.center)]
        \begin{feynman}
          \node [dot] (a);
            \node [left=1.5cmof a] (b);
            \node [right=0.8 cmof b] (bb); 
            \node [above = 0.35cmof bb] (name1) {\large{$N_i$}};
            \node [above right=1.75cmof a] [dot] (c) ;
            \node [right= 0.35 cmof a] (cc);
            \node [above=1 cmof cc] (name) {\large{$H_u$}};
            \node [below right=1.75cmof a] [dot] (d);
            \node [right=0.35 cmof a] (dd);
            \node [below=1 cmof dd] (namel) {\large{$l$}};
            \node [right = 1.8 cmof c] (e);
            \node [right=1cmof c] (ee);
            \node [above=0.35cmof ee] (eee) {\large{$l_\alpha$}};
            \node [right =1.8cmof d] (f);
            \node [right=1 cmof d] (ff);
            \node [below=0.35 cmof ff] (fff) {\large{$H_u$}};
            \node [right=1.6 cmof a] (aa) {\large{$N_j$}};
            \diagram* {
                (b) -- [plain,thick] (a);
                (a) -- [scalar,thick] (c),
                (d) -- [fermion,thick] (a),
                (c) -- [fermion,thick] (e),
                (d) -- [scalar,thick] (f),
                (c) -- [plain,thick] (d)
            };
            \end{feynman}
    \end{tikzpicture}
    \begin{tikzpicture}[baseline=(current bounding box.center)]
        \begin{feynman}
          \node [dot] (a);
            \node [left=1.7cmof a] (b);
            \node [right=0.85cmof b] (bb); 
            \node [above = 0.35cmof bb] (name1) {\large{$N_i$}};
            \node [right = 1.7cmof a] [dot] (c);
            \node [right=0.85 cmof a ] (cc);
            \node [below=1.2 cmof cc] (name) {\large{$l,\bar{l}$}};
            \node [above=1.2 cmof cc] (name2) {\large{$H_u$}};
            \node [right=1.2cmof c] [dot] (d);
            \node [right=0.6cmof c] (dd);
            \node[ above=0.35 cmof dd] (name3) {\large{$N_j$}};
            \node [above right=1.7cmof d] (f){\large{$H_u$}};
            \node [below right=1.7cmof d] (g){\large{$l_\alpha$}};
            \diagram* {
                (b) -- [plain,thick] (a),
                (a) -- [plain,half right,thick] (c),
                (c) -- [scalar,half right,thick] (a),
                (c) -- [plain,thick] (d),
                (d) -- [scalar,thick] (f),
                (d) -- [fermion,thick] (g)
            };
            \end{feynman}
    \end{tikzpicture}
    \caption{\it CP-violating $N_i$ decay. }\label{feynman}
    \end{figure}
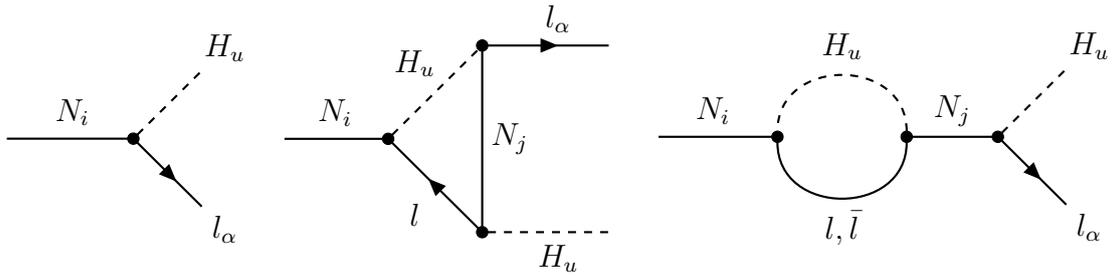

The lepton number violating decays of the sterile Majorana neutrinos $N_i$ are the main ingredients for the thermal leptogenesis recipe \cite{Buchm_ller_2004,GIUDICE200489}. They depend, in general, on the mass of the heavy neutrinos, as well as their Yukawa couplings with the charged leptons and the Higgs doublet. In order to generate a lepton asymmetry, we first need a thermal production of a distribution of $N_i$ through scattering in the early Universe at temperatures around $T\sim M_{N_i}$, where we indicate with $M_{N_i}$ the mass of the $N_i$ sterile neutrino.
As the temperature drops below the mass $M_{N_i}$, these sterile neutrinos undergo decays via processes that violate the lepton number by one unit ($\Delta L=1$),  
shown in figure \ref{feynman} and, if these decays are slow enough compared to the expansion rate of the Universe, the $N_i$ abundance does not decrease according to the Boltzmann equilibrium statistics $\propto e^{-M_{N_i}/T}$, as the equilibrium demanded. This is how a net lepton asymmetry is generated. Thus, the crucial ingredient is to estimate the mass of the heavy neutrinos. To this aim, it is useful to recast the Majorana and Dirac mass matrices of eqs. \eqref{mdirac} and \eqref{mmajo} in the following way:
\begin{equation}
    M_D=\;v_u \,g  Y_1^2\,\begin{pmatrix}\left( 1+g^{\prime\prime}/g\right)+\left( g^{\prime\prime}/g-1\right)\zeta^2&2\,\zeta&2\,\zeta\,g^\prime/g\\
    2\,\zeta&\left( 1+g^{\prime\prime}/g\right)\zeta+\left( g^{\prime\prime}/g-1\right)&(1-\zeta^2)\,g^\prime/g
    \end{pmatrix}_{\text{RL}}\;,
    \label{eq:DiracMass}
    \end{equation}
    \begin{equation}
    \mathcal{M}_R=\;\Lambda\,Y_1^2\,\begin{pmatrix}\lambda (1+\zeta^2)-(\zeta^2-1)&2\,\zeta\\
    2\,\zeta&\lambda (\zeta^2+1)+(\zeta^2-1)
    \end{pmatrix}_{\text{RR}}\;,
\label{eq:MajoranaMass}
\end{equation}
where $\zeta=Y_2/Y_1$. With the modulus $\tau$ in the fundamental domain, and for small values of $\text{Re}\,\tau$, it reads:
\begin{equation}
    \zeta\simeq 8\,\sqrt{3}\,t^2\,\left( 1+i\,\pi\,\text{Re}\tau\right)\,,\quad t\equiv e^{-(\pi\,\text{Im}\tau)/2}\,,
\end{equation}
where only terms up to $\mathcal{O}(t^2)$ are considered. \\
Therefore, with $|\zeta|\ll 1$ and all the free parameters close to their best fit values quoted in table \ref{seesaw_table}, it is evident that the mass scale of the right-handed neutrinos is basically $\mathcal{M}_N\sim \Lambda Y_1^2$ and the Yukawas $\mathcal{Y}\sim g\,Y_1^2$. In other words, both the mass scale of the Majorana neutrinos and the Yukawas get suppressed by the modular form $Y_1$, which can be expanded as:
\begin{equation}
    Y_1(\tau)\simeq \dfrac{7}{100}+\dfrac{42}{25}\,t^2\,\left( 1+2\,i\,\pi\,\text{Re}\tau\right)+\mathcal{O}(t^3)\,.
\end{equation}
Since the mass-scale of light neutrinos is completely determined by the global parameter $g^2\,v_u^2/\Lambda$, we can estimate $\Lambda$. 
By imposing that the dimensionless parameter $g$ assumes a value of order $\mathcal{O}(1)$, we get $\Lambda \sim \mathcal{O}(10^{14})$ GeV. 
This, along with the suppression provided by the modular form $Y_1$, implies a mass scale $\mathcal{M}_N$ for the Majorana neutrinos of approximately $10^{12}$ GeV. 

In principle, the interference between the tree-level and the loop contributions in figure \ref{feynman} produces CP violating decays that can generate asymmetries in all flavors $\alpha$. 
Flavor effects \cite{Endoh_2004,Abada_2006,Nardi_2006} become relevant when the rate of interaction $\Gamma_\beta$ of the processes involving a lepton with flavor $\beta$ ($\beta=e,\mu,\tau$) are rapid enough to discern distinct lepton flavors. Under these circumstances, the coherence of leptons and antileptons as flavor superpositions is lost during leptogenesis. Consequently, it becomes necessary to adopt a flavor basis to calculate the baryon asymmetry.
Typically, the temperature below which the flavor effect cannot be neglected is such that the interaction rate $\Gamma_\beta$ for a charged lepton with flavor $\beta$ becomes faster than the expansion rate $H(T)$ of the Universe.
The interaction rate can be estimated \cite{Campbell_1992,Cline_1994} as:
\begin{equation}
    \Gamma_\beta \simeq 10^{-2}\, \alpha_\beta^2\, T\;,
\end{equation}
where $\alpha_\beta$ is the charged-lepton Yukawa coupling.
For a certain flavor-dependent $T\lesssim T_\beta$, the condition $\Gamma_\beta (T)>H(T)$ is satisfied, so that the typical temperatures below which the leptogenesis becomes sensitive to lepton flavors are $T_e\simeq 4\cdot 10^{4}$ GeV, $T_\mu\simeq 2\cdot 10^{9}$ GeV and $T_\tau\simeq 5\cdot 10^{11}$ GeV \cite{Davidson:2002qv,Covi_1996,Hamaguchi_2002}. The amount of CP asymmetry generated by the interference between the tree and one loop decay diagrams of figure \ref{feynman}
can be described by the $\epsilon_i$ parameters, defined as:
\begin{equation}
\begin{aligned}
    \epsilon_i\,=&\sum_\alpha\dfrac{\Gamma(N_i\rightarrow L_\alpha\,H_u)-\Gamma(N_i\rightarrow \overline{L_\alpha}\,H_u^\ast)}{\Gamma(N_i\rightarrow L_\alpha\,H_u)+\Gamma(N_i\rightarrow \overline{L_\alpha}\,H_u^\ast)}=\\
    =&\,\sum_{j\not=i}\dfrac{\text{Im}\left[\left( \mathcal{Y}^\dagger \mathcal{Y}\right)^2_{ji}\right]}{8\pi(\mathcal{Y}^\dagger \mathcal{Y})_{ii}}\left[ f(x_{ij})+g(x_{ij})\right]\,\quad \text{with} \;i=1,2
    \end{aligned}
    \label{eq:efficiencyandCPV}
\end{equation}
with $x_{ij}=M_j^2/M_i^2$ and the loop functions \cite{Pascoli_2007,Branco_2009}:
\begin{equation}
    \begin{aligned}
        f(x_{ij})\,=\,&\sqrt{x_{ij}}\Big[ 1-(1+x_{ij})\log\left( 1+\frac{1}{x_{ij}}\right)+\frac{1}{1-x_{ij}}\Big]\,,\\
        g(x_{ij})\,=\,&\dfrac{\sqrt{x_{ij}}(1-x_{ij})}{(1-x_{ij})^2+x_{ij}(\mathcal{Y}^\dagger \mathcal{Y})^2_{jj}/(16\pi^2)}\;,
        \label{loopsFunc}
    \end{aligned}
\end{equation}
which are related to the vertex correction and the self energy correction, respectively.\\
The lepton number violation produced by the sterile neutrino decays can be depleted by scatterings and inverse processes, which are often referred to as \emph{wash-out processes}. If they occur out-of-equilibrium, \emph{i.e.} they are not very fast compared to the Hubble expansion rate of the Universe $H(T)$ at temperature $T=M_{N_i}$, a net asymmetry can survive. 
\\
Typically, we can quantify the departure from thermal equilibrium by the \emph{decay parameter} which, in terms of  the decay rates $\Gamma_i$  of the $N_i$ sterile neutrino, reads as follows:
\begin{equation}
    K_i\equiv \dfrac{\Gamma_i}{H(T=M_i)}=\dfrac{\widetilde{m_i}}{m^\ast}\;,
\end{equation}
where the \emph{effective neutrino mass} \cite{Buchm_ller_1996} is defined as:
\begin{equation}
    \widetilde{m_i}=\dfrac{\left(\mathcal{Y}^\dagger \mathcal{Y}\right)_{ii}\,v_u^2}{M_i}\;,
\end{equation}
and $m^\ast \simeq 1.1\cdot 10^{-3}$ eV.
We can distinguish three different regimes: $K\ll 1$, $K\simeq 1$ and $K\gg 1$, which we identify as the \emph{weak}, \emph{intermediate} and \emph{strong} wash-out regime, respectively. Using our best fit values of the free parameters of the model, we deduce we are always in a strong wash-out regime; thus, the contributions to the leptogenesis 
from the $\Delta L=1$ and $\Delta L=2$ processes\footnote{
We refer to the relevant lepton number violating processes in the thermal plasma: the $\Delta L=1$ scatterings between $N_i$ and leptons mediated by the Higgs boson, involving the top quark or electroweak gauge bosons, and the $\Delta L=2$ scatterings between the leptons and the Higgs boson with an intermediate $N_i$ heavy neutrino. As discussed in \cite{Pramanick:2024gvu}, in some hybrid leptogenesis frameworks such scattering processes must be necessarily considered since they can significantly alter the final amount of baryon asymmetry.} can be safely neglected, as discussed in \cite{Davidson_2008,Buchm_ller_2003,Buchmuller:2002rq}. Therefore, from now on, we will only consider the decays and the inverse decays.
Let us now discuss the relevant Boltzmann Equation (BE) for the leptogenesis in our model, with two heavy sterile neutrinos $N_{1,2}$ with masses $M_1\lesssim M_2$. Assuming the reheating temperature $T_\text{rh}>M_{1,2}$, both the sterile neutrinos $N_{1,2}$ are thermally produced \cite{croon2019stability,GIUDICE200489,Giudice_1999,Khlopov:1984pf} in the Early Universe, 
and the set of classical kinetic equations \cite{Buchmuller:2004nz} could be written as:\footnote{It is convenient to write the BE for the $B-L$ asymmetry evolution, because all the electroweak processes conserve it, including those mediated by the sphalerons.}
\begin{equation}
    \begin{aligned}
        \dfrac{\text{d} N_{N_i}}{\text{d} z}=&-D_i\left( N_{N_i}-N_{N_i}^{eq}\right)\,,\quad\text{with } i=1,2\\
        \dfrac{\text{d} N_{B-L}}{\text{d} z}=&-\sum_{i=1}^2\epsilon_i\,D_i\left( N_{N_i}-N_{N_i}^{eq}\right)-\sum_{i=1}^2W_i\,N_{B-L}
    \end{aligned}
    \label{eq: BE}
\end{equation}
with $z=M_1/T$. Here, $D_i$ is the decay term while $W_i$ is the washout factor; futhermore,
the $N_i$'s are the  number densities of the RH sterile neutrinos, while $N_{B-L}$ is the amount of $B-L$ asymmetry, both normalized by the comoving volume \cite{Buchmuller:2004nz}. With this choice, 
we automatically take into account the expansion of the Universe, which was not possible if we were to employ the number density $n_X$ of the particle species.
The equilibrium abundances of $N_i$'s are given by $N_i^{eq}=\frac{3}{8}z_i^2\mathcal{K}_2(z_i)$, with $\mathcal{K}_2$ the Bessel function of the second kind. In the literature, the comoving volume $R^3_\ast(t)$ is usually chosen as to contain one photon at the time $t_\ast$, before the onset of the leptogenesis \cite{Buchmuller:2002rq}:
    \begin{equation}
        N_X(t)=n_X(t)R^3_\ast(t)\, ,
    \end{equation}
    with 
    \begin{equation}
        R_\ast(t_\ast)=\left(n_\gamma^{eq}(t_\ast) \right)^{-1/3}\, ,
    \end{equation}
    so that $N_\gamma(t_\ast)=1$.
    Using $z_i=z\sqrt{x_{1i}}$, we can write the decay term $D_i$ and the washout factor $W_i$ as:
    \begin{equation}
        \begin{aligned}
            D_i=&\,K_i\, x_{1i}\, z\, \braket{1/\gamma_i}\,,\\
           W_i\simeq W_i^{ID}=&\,\dfrac{1}{4}\,K_i\sqrt{x_{1i}}\mathcal{K}_1(z_i)z_i^3\,,
        \end{aligned}
        \label{eq:DecayAndWashout}
    \end{equation}
    where $W_i^{ID}$ is the inverse decay term which contributes to the washout of the lepton asymmetry, and $\braket{1/\gamma_i}$ is the thermally averaged dilation factor, given as the ratio of the modified Bessel functions of the second kind:
    \begin{equation}
        \Biggl \langle\dfrac{1}{\gamma_i}\Biggr \rangle=\dfrac{\mathcal{K}_1(z_i)}{\mathcal{K}_2(z_i)}\,.
    \end{equation}
    The final $B-L$ asymmetry $N_{B-L}^f$ is given by:
    \begin{equation}
        N_{B-L}^f=N_{B-L}(z=\infty)=N_{B-L}^{in}e^{-\sum_i\int dz^\prime W_i (z^\prime)}+N_{B-L}^{lep}\, ,
    \end{equation}
    where $N_{B-L}^{lep}$ is the $B-L$ produced via thermal leptogenesis and $N_{B-L}^{in}$ indicates the possible pre-existing asymmetry \cite{Alexander:2006lty,Adshead:2017znw}, at an initial temperature $T_{in}$ before the leptogenesis onset. The potential pre-existing $B-L$ asymmetry could introduce further constraints on the parameter space of the model \cite{Giudice_1999,Bertuzzo_2011,Di_Bari_2013,Chianese_2018}. In order to consider a scenario of pure leptogenesis from RH neutrino decays, we will fix $N_{B-L}^{in}=0$ throughout the rest of the paper.
    
    During the expansion of the Universe, the $B-L$ asymmetry is reprocessed in a baryon asymmetry through sphaleron processes. If sphalerons decouple before the electroweak phase transition, the amount of $B$ asymmetry produced is \cite{PhysRevD.42.3344}:
\begin{equation}
    N^f_{B}=\dfrac{(8\,N_F+4\,N_H)}{(22\,N_F+13\,N_H)}N^f_{B-L}\;,
\end{equation}
where $N_F=3$ and $N_H=2$ stand for the number of fermion generations and the number of Higgs doublets, respectively. The predicted baryon-to-photon ratio has to be compared with the value $\eta_B$ measured at recombination \cite{2020}. The relation is:
\begin{equation}
    \eta_B=\dfrac{N_{B}^f}{f} \,,
\end{equation}
where $f=2387/86$ is the dilution factor calculated assuming standard photon production from the onset of leptogenesis till recombination \cite{Buchmuller:2004nz}.
\subsection{Validity of the $N_1-$dominated scenario} \label{subsec:validity}
\begin{figure}[t]
    \centering
    \includegraphics[width=1.05\textwidth]{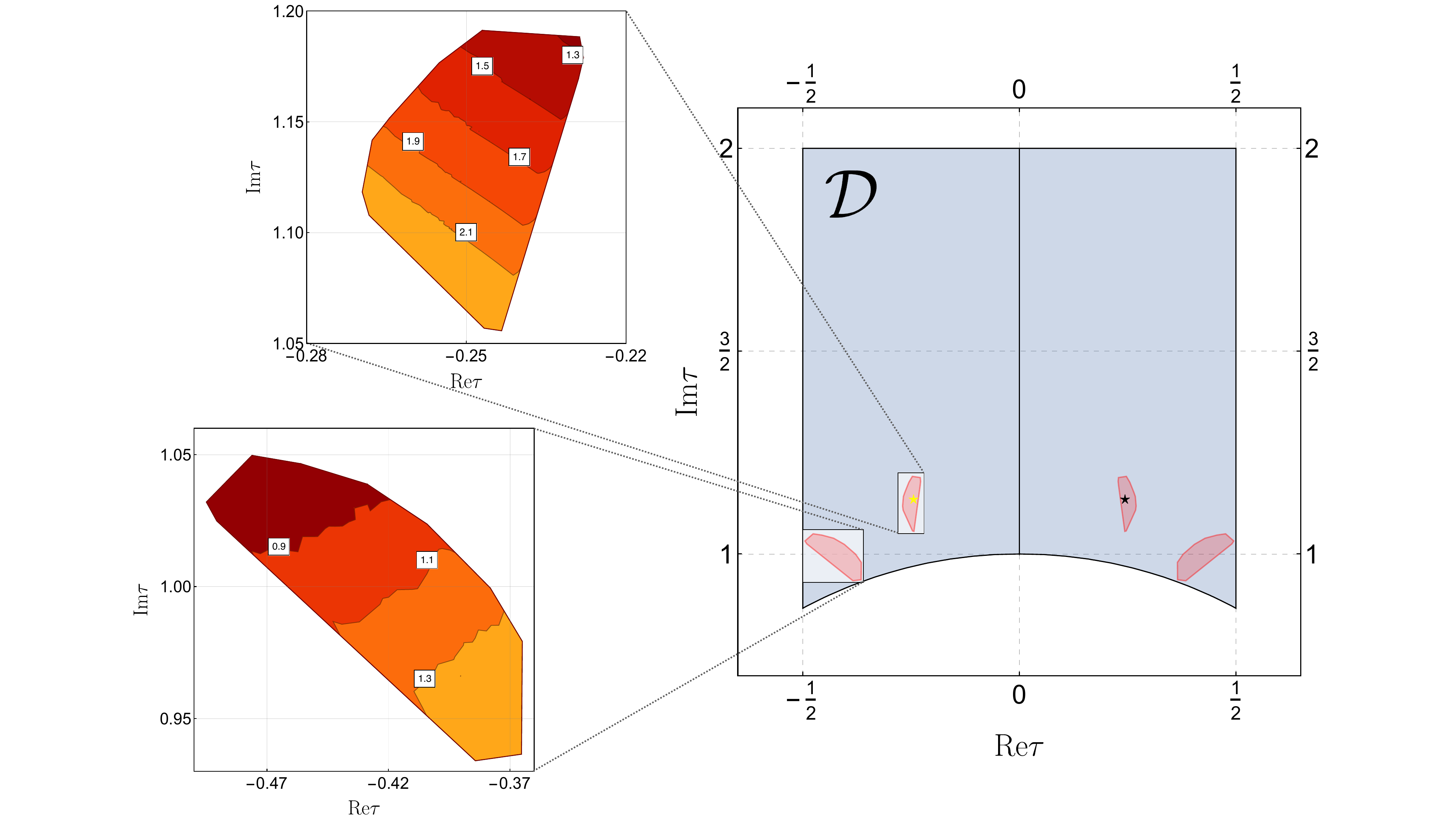}
    \caption{\textit{Contour plots showing the values assumed by $\delta_2$ in the best fit regions. For a clearer and more concise data visualization, all the fitted points from section \ref{numres} were mapped into $\mathcal{D}$, see appendix \ref{rescD} for details.}}
    \label{fig:zoom}
\end{figure}

In presence of a hierarchical mass spectrum $M_1\lesssim M_2$, only the contribution from the lightest sterile neutrino to the leptogenesis is commonly considered, the so-called $N_1-$dominated scenario. However, as widely discussed in literature \cite{Samanta_2020,Blanchet_2012,Blanchet_2006}, there exists a parameter space for which such an assumption is no longer valid, and the heaviest sterile neutrino contribution to the final baryon asymmetry cannot be neglected.
An important parameter that can be used to test whether the $N_1-$dominated scenario sets in  is the parameter $\delta_i=(M_i-M_1)/M_1$. This is related to $x_{1i}$ by $ \sqrt{x_{1i}}=1+\delta_i$, and it will tell us whether the CP-asymmetry can undergo resonant enhancement \cite{Davidson_2008,Pilaftsis:2003gt,Dev:2017wwc,Pilaftsis_2005,Xing_2007}. In fact, it can be noted that, when dealing with an almost degenerate mass spectrum, the self-energy loop function $g(x_{ij})$ in eq.\eqref{loopsFunc} can be reformulated in terms of $\delta_j$ in the following manner:
\begin{equation}
   g(\delta_j)\simeq\,\dfrac{\delta_j}{2\left[(\delta_j)^2+(\mathcal{Y}^\dagger \mathcal{Y})^2_{jj}/(64\pi^2)\right]}\;.
\end{equation}
Therefore, if the Yukawa matrix and the mass splittings are independent quantities, the resonance enhancement occurs if the condition \cite{Pascoli_2007}
\begin{equation}
    \delta_j^2\simeq (\mathcal{Y}^\dagger \mathcal{Y})^2_{jj}/(64\pi^2)
    \label{eq:deltaSq}
\end{equation}
holds. As already noted, our model provides Yukawa couplings suppressed by the modular form $Y_1$ while the mass splitting parameter $\delta_2$ belongs to the range $\left[ 0.7,2.3\right]$, as our evaluation shows in figure \ref{fig:zoom}.
Given that the right-hand side of eq.\eqref{eq:deltaSq} is approximately on the order of $\mathcal{O}(10^{-5})$, we safely deduce that the resonance condition is never satisfied in the best fit regions.\footnote{Notice that the resonance would have probably ruined the prediction of the model regarding the baryon-to-photon ratio $\eta_B$; indeed, with heavy sterile neutrinos with masses around $10^{12}$ GeV, the resonance would imply an enhancement of the CP-violating parameter such that the final baryon asymmetry exceeds the observed value by many order of magnitudes (see \emph{e.g.} \cite{Arcadi_2023}).}
We can now discuss the validity of the $N_1-$dominated scenario in presence of another heavy neutrino $N_2$. In general, one should solve eqs.\eqref{eq: BE} taking into account the wash-out effects due to both $N_1$ and $N_2$ sterile neutrinos. The strength of the $N_2$ wash-out effects is encoded into $W_2^{ID}$, see eq.\eqref{eq:DecayAndWashout}, which depends on the $\delta_2$ parameter. 
There exists a specific value  $\delta_2^{\ast}$ above which the influence of $W_2^{ID}$ becomes irrelevant compared to $W_1^{ID}$; in that case, leptogenesis would be predominantly driven by the presence of the lightest neutrino, from which the name of $N_1-$dominated scenario. 
Conversely, for $\delta_2<\delta_2^\ast$, the effect of $W_2^{ID}$  becomes relevant. 
Usually, in a strong wash-out regime, a value $\delta_2^\ast\gtrsim1.5\div 5$ is found \cite{Samanta_2020,Blanchet_2006}.
In our model, $\delta_2^\ast\simeq 1$, so that the $N_1-$dominated scenario is valid for $\text{Re}\tau\in[-0.44,-0.22]$ (which includes our best fit point) where $\delta_2 \gtrsim \delta_2^\ast$, while in the other regions of the parameter space we cannot a priori neglect the wash-out effects of the $N_2$ sterile neutrinos. In appendix \ref{sec:N1dominato} we illustrate that in the absence of specific correlations among the parameters, the contribution of the heavier sterile neutrino to leptogenesis is negligible for $\delta_2\gtrsim 1.5$.
\newpage
\subsection{Numerical analysis}\label{lepto_analysis}
\begin{figure}[t]
    \centering
    \includegraphics[width=0.8\textwidth]{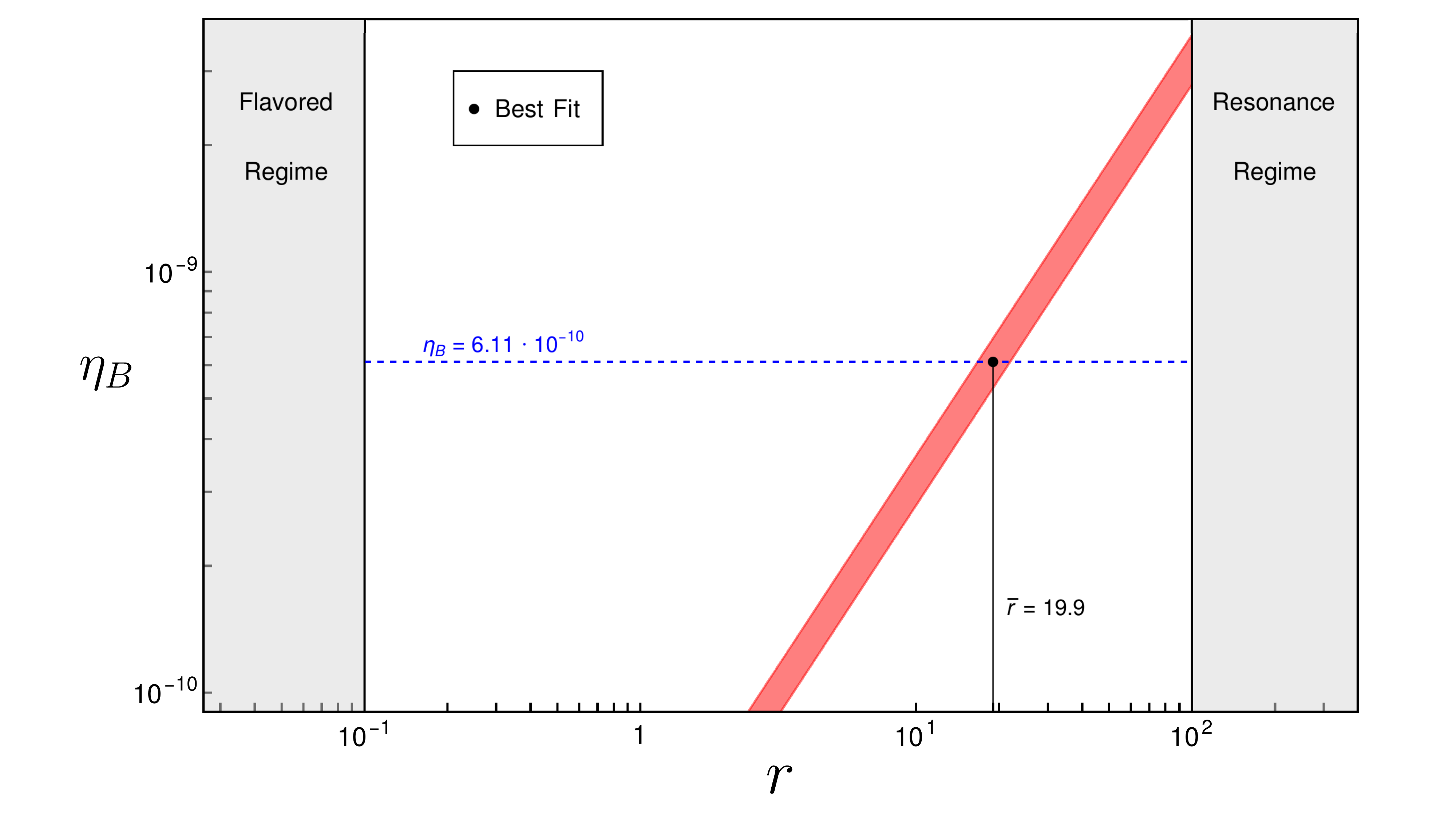}
    \caption{\textit{Pink band: dependence of the final baryon-to-photon asymmetry $\eta_B$ on the rescaling parameter “$r$" ($\Lambda\to r\,\Lambda$) for a fixed set of the theory's free parameters, satisfying $\sqrt{\chi^2}\leq 2$. The observed value of $\eta_B$ \cite{2020} is shown with the blue line. The unflavored leptogenesis is provided for $10^{-1}<r<10^{2}$, and it turns out to be successful for $16.8\leq r\leq22.1$. In particular, at the best fit, $\eta_B$ in agreement with the observation for $r=\bar{r}=19.9$.}}
    \label{fig:scanR}
\end{figure}
We solve the full Boltzmann Equations presented in eqs.\eqref{eq: BE} for the best fit points that satisfy $\sqrt{\chi^2}\leq 2$. With the heavy sterile neutrino masses at $\mathcal{O}(10^{12})$ GeV, the model provides $\eta_B\lesssim 10^{-10}$, slightly below the observed value. 
However,  since only the ratio $g^2\, v_u^2/\Lambda$ is constrained by the masses of the light neutrinos, the mass scale of the Majorana neutrinos can be regarded as a free parameter of the model. We can safely rescale $\Lambda\rightarrow r\,\Lambda$ and $g\rightarrow \sqrt{r}\,g$ without altering the results of the fit in table \ref{seesaw_table}.\footnote{A similar work has been done in \cite{Jung_2022}, in a context of a minimal extended seesaw, with an additional singlet field.} Notice that varying the mass-scale of the right-handed neutrinos have relevant consequences. For $r\lesssim 10^{-1}$ the mass-scale of the sterile neutrinos is lowered down to $\mathcal{O}(10^{11})$, below which the \emph{flavored regime} sets in; on the other hand, if $r\gtrsim 10^2$ the \emph{resonant regime} occurs. We confine the analysis to the range $r\in \left[ 10^{-1},10^{2}\right]$, focusing on the unflavored scenario that involves solely the lightest sterile neutrino \cite{Fukugita:1986hr,Buchmuller:2004nz,Barbieri:1999ma,Buchmuller:2002rq,Buchm_ller_2003,GIUDICE200489,Blanchet:2008pw}. The results are shown in figure \ref{fig:scanR}, where the pink band represents the dependence of the final baryon-to-photon asymmetry $\eta_B$ on the parameter $r$ for a fixed set of the theory's free parameters, satisfying $\sqrt{\chi^2}\leq 2$; in \emph{blue}, instead, we report the observed value of $\eta_B$ (due to the small uncertainty on this observation, the $1\sigma$ allowed range is basically indistinguishable from the shown dotted line).
In particular, the best fit realization of our model provides a successful leptogenesis if $\Lambda\rightarrow \bar{r}\Lambda$, where $\bar{r}=19.9$. 
It is important to observe that this selection does not compromise the quality of our fit; in fact, it modifies $g$ by a multiplicative factor of approximately $\sqrt{\bar{r}}\sim 4$, thereby maintaining $g$ as an order $\mathcal{O}(1)$ parameter, as for most of the other parameters of the model.


\section{Conclusions}
\label{concl}
In the present work, we have addressed the problem of finding a modular invariant minimalistic construction based on the smallest modular finite group $\Gamma_2\cong S_3$ that was compatible with low energy neutrino oscillation data and, with the same model parameters, able to accommodate the measured value of the baryon asymmetry of the Universe (BAU). Our attempts have been guided by the requirement to minimally extend the Standard Model particle content; with the aim to naturally generate small neutrino masses, we relied on a seesaw mechanism built with the help of only two heavy right-handed neutrinos. A straightforward prediction of our model is that the lightest neutrino mass is zero at tree level. Having imposed a generalized CP-symmetry, not only we were able to restrict the number of free parameters to ten (eight dimensionless, of which two come from $\tau=\re \tau+i\,\im \tau$, and two global scales) but we have also secured that the modulus $\tau$ is the only source of CP-violation. We showed that, even within this limitation,  this was enough to reproduce the current experimental value of the leptonic Jarlskog invariant within $1\sigma$. Furthermore, the model strongly favours the Normal Ordering of neutrino masses, a prediction that can be tested in upcoming experiments \cite{DUNE:2020jqi, Hyper-KamiokandeProto-:2015xww, JUNO:2015zny,IceCube-Gen2:2019fet}. The predictions for the Majorana effective mass of the neutrinoless double-beta decay and the effective neutrino mass of tritium decay are both substantially within the current upper bounds from \cite{KamLAND-Zen:2022tow, KATRIN:2021uub}, and are both analitically understood from the constraint on the lightest neutrino mass, which is in turn a consequence of the Minimal seesaw scenario.

Since the only source of CP violation in our model is dictated by $\re \tau$, it was highly non-trivial to generate a sufficient amount of BAU through the leptonic asymmetry $\eta_B$; we have demonstrated, indeed, that this is the case in our model.  Given that in the presence of a mild hierarchical heavy mass spectrum with $\delta_2\,\gtrsim \,   1$ only the contribution from the lightest
sterile neutrino to the leptogenesis is commonly considered, we solved the appropriate Boltzmann Equations in the  $N_1$-dominated scenario and shown that a mild rescaling of the Majorana right-handed mass
scale $\Lambda$ (not fixed by the fit) is enough to guarantee the required $\eta_B$ value. Given the minimality of our $\Gamma_2\cong S_3$ construction, we consider this feature a relevant result in the model building based on modular invariant theories. 


\acknowledgments
The authors would like to thank J. Penedo for useful discussions.


\appendix

\section{Beyond the $N_1-$dominated scenario}
\label{sec:N1dominato}
\renewcommand{\thefigure}{\Alph{section}.\arabic{figure}}
\setcounter{figure}{0}
\begin{figure}[H]
    \centering
    \includegraphics[width=.495\textwidth]{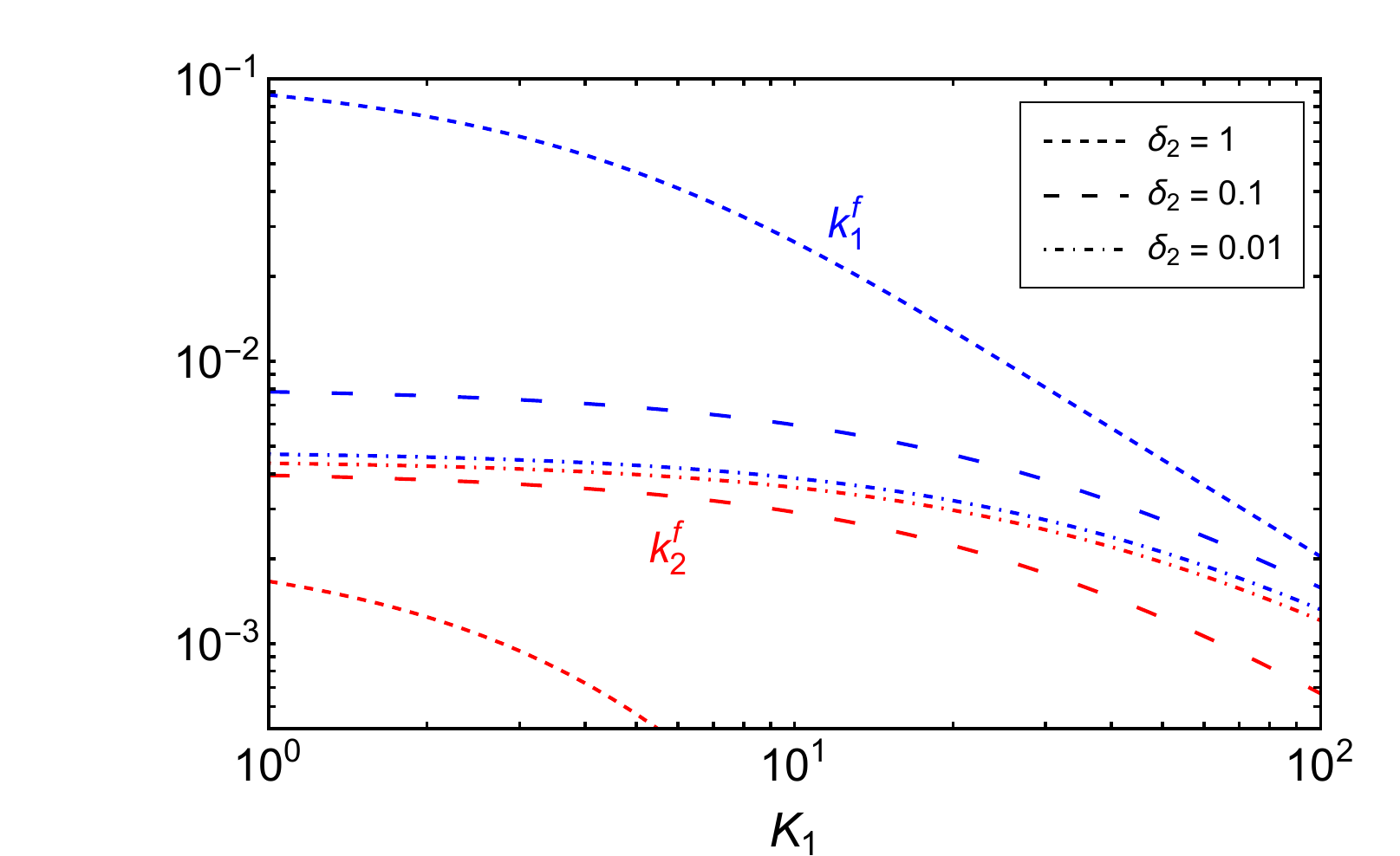}
    \includegraphics[width=.495\textwidth]{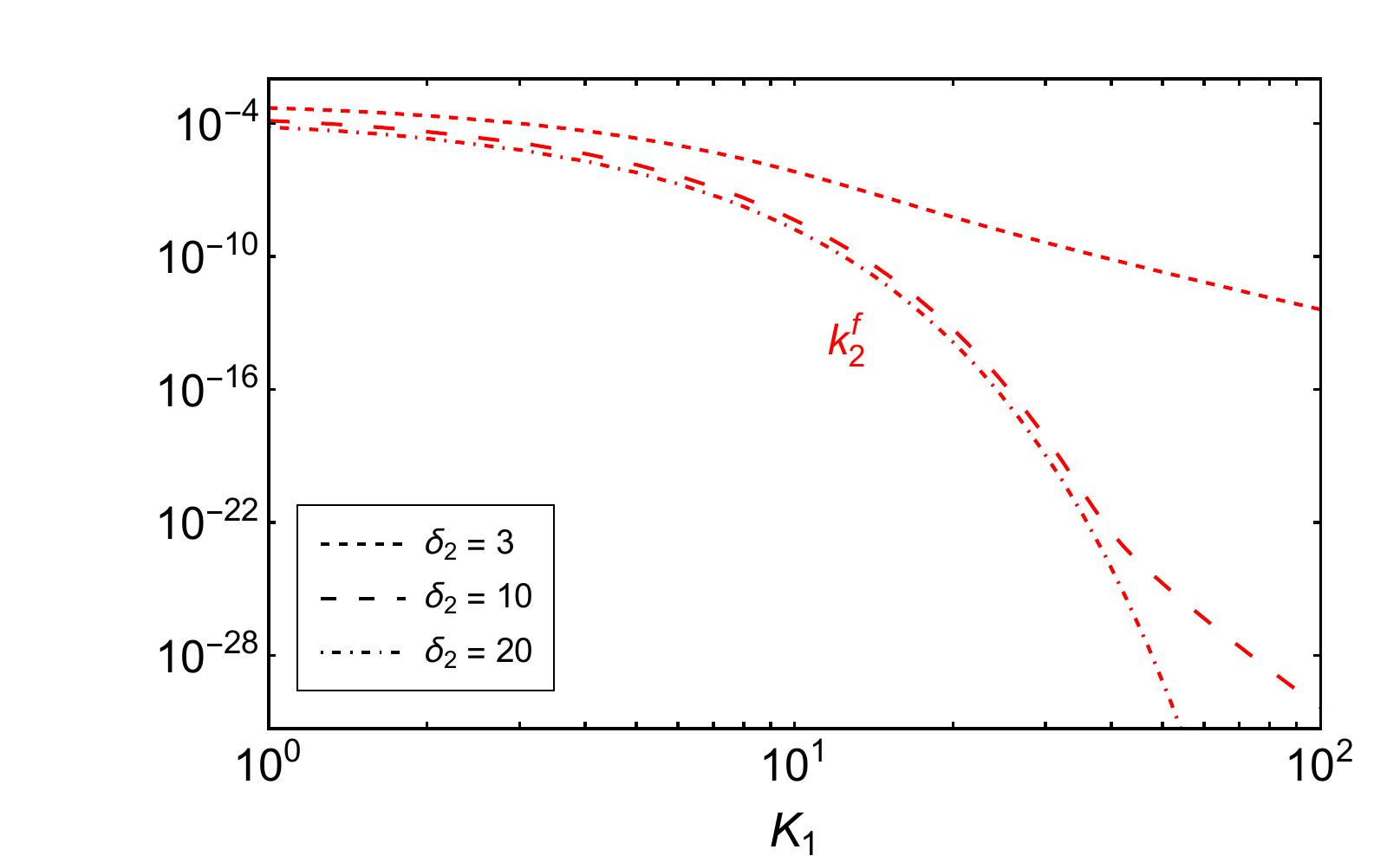}\\
    \includegraphics[width=.495\textwidth]{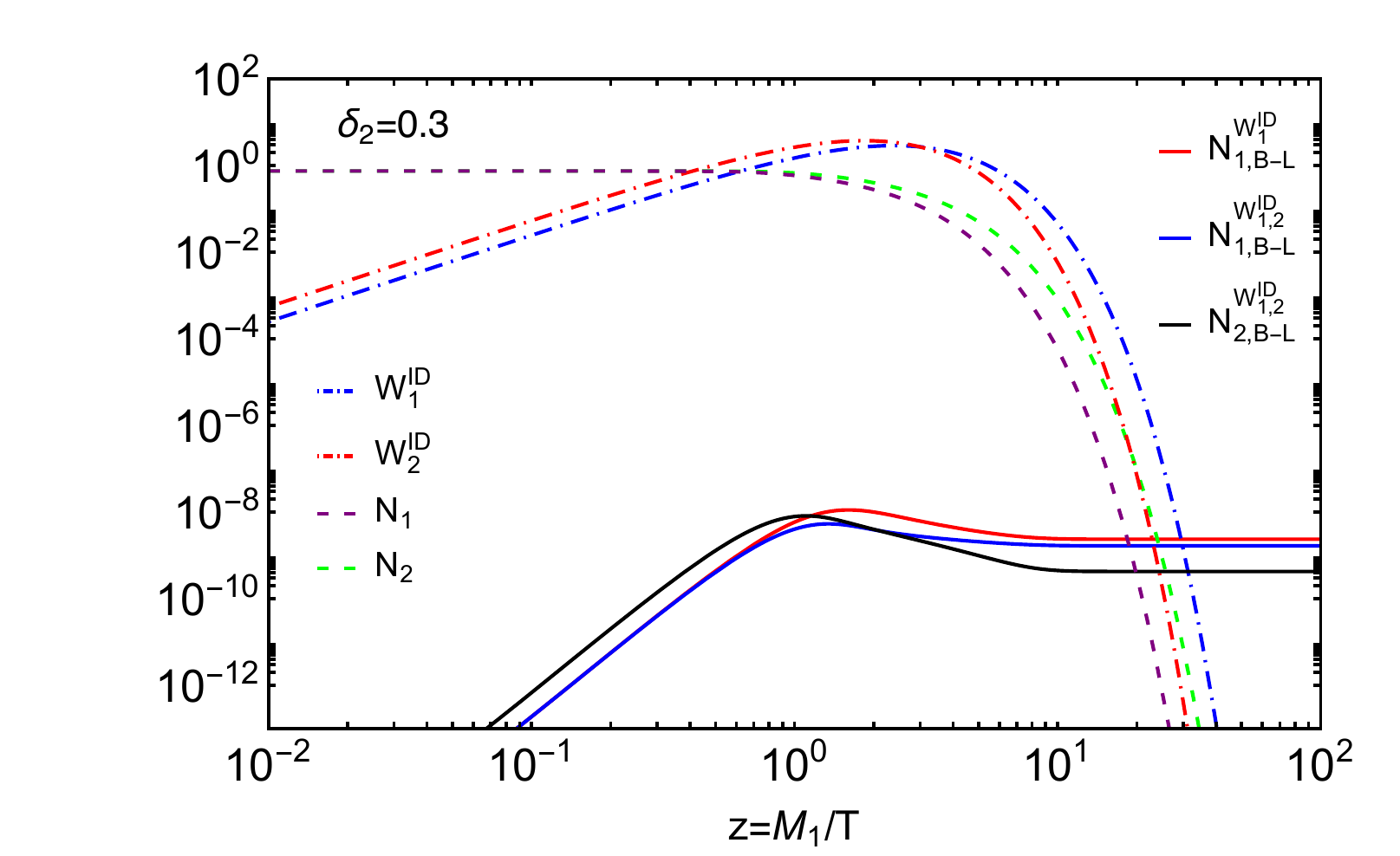}
    \includegraphics[width=.495\textwidth]{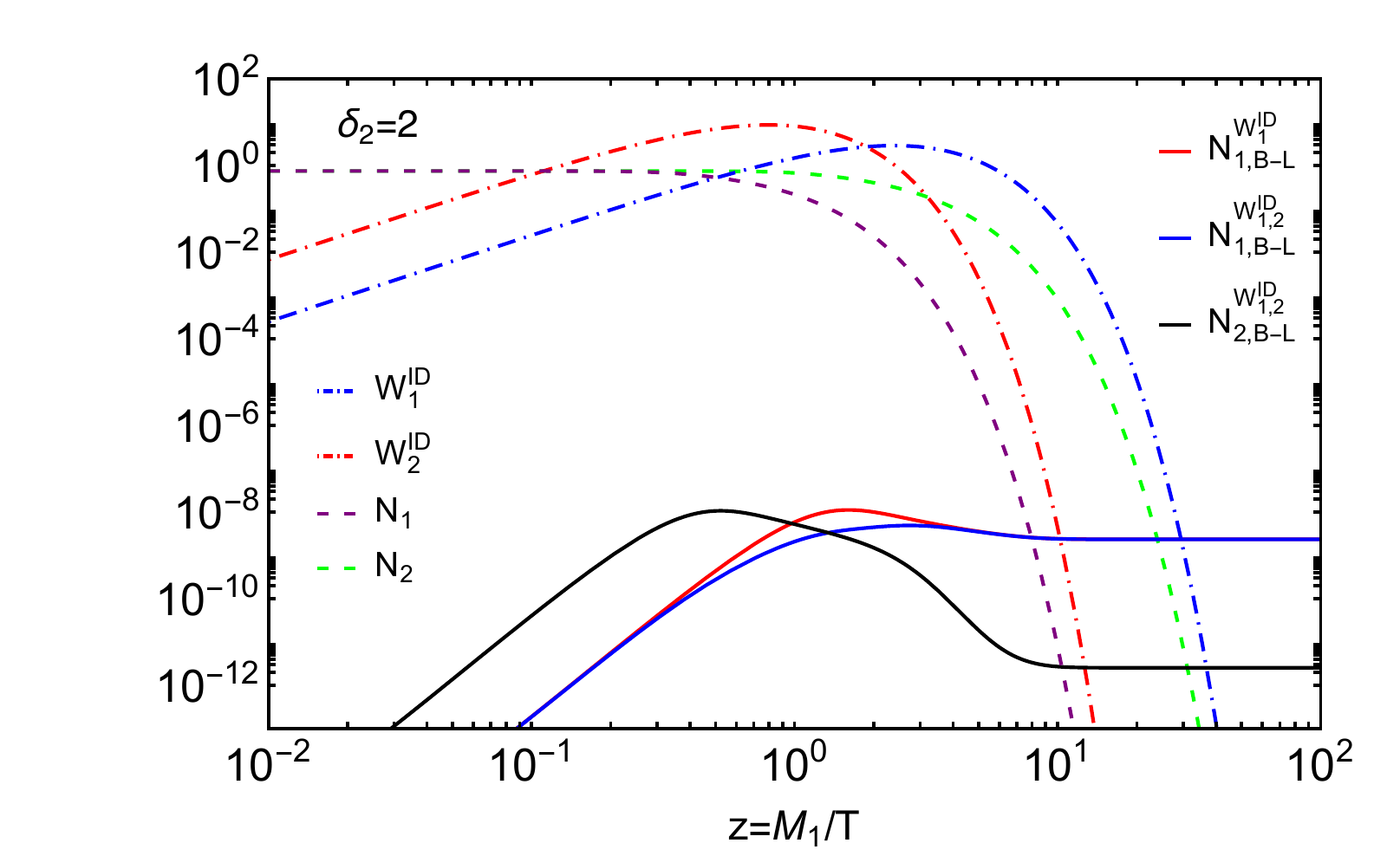}
    \caption{\textit{
    Top left (right) panel: evolutions of the efficiency factors $k_i^f$ in terms of $K_1$ with $\delta_2=1,\,0.1,\,0.01$ ($\delta_2=3,\,10,\,20$), with $K_2\simeq 
    25$. The efficiency factors of $N_1$ and $N_2$ are shown in blue and red dashed lines, respectively.  Bottom left (right) panel: $|N_{B-L}|$ as a function of $z=M_1/T$ and other relevant quantities such as $N_i$ and $W_i^{ID}$, with $\delta_2=0.3$ ($\delta_2=2$), $N_i^{in}=3/4$ and $K_1=K_2=10$. 
    The colour code for $N_i, W_{i}^{ID}$ is highlighted in the legend. Solid
red line shows $|N_{B-L}|$ for a pure $N_1-$dominated scenario. The solid blue line shows the asymmetry
generated by $N_1$, partially depleted by the wash-out effects of $N_1$ and $N_2$, given by $W_{1}^{ID}$ and $W_{2}^{ID}$,
respectively.
The solid black line indicates $|N_{B-L}|$ produced by $N_2$, undergoing $W_{1,2}^{ID}$ washout.}
    }
    \label{fig:BeyondN1DS}
\end{figure}
In this appendix we firstly illustrate that, in the absence of specific correlations among the decay parameters, the feasibility of the $ N_1$-dominated scenario hinges crucially on the value of the $\delta_2$ parameter, defined in section \ref{subsec:validity}. This is shown in figure \ref{fig:BeyondN1DS}. Then, we discuss the concrete example related to the best fit realization of our model.\\
In presence of two right-handed sterile neutrinos, the solution of the BEs in eq.\eqref{eq: BE} can be written as \cite{Kolb:1990vq}: 
\begin{equation}
    N_{B-L}^f=\sum_{i} \epsilon_i k^f_i\,\,,\quad i=1,2
    \label{eq:NBLAnalitica}
\end{equation}
with $\epsilon_i$ the usual CP-violating parameters (see eqs.\eqref{eq:efficiencyandCPV} and \eqref{loopsFunc}) and $k_i^f$ the \emph{efficiency factors}. Without strongly hierarchical Yukawa couplings, for quasi-degenerate heavy neutrinos we can assume $\epsilon_1\simeq \epsilon_2$. Thus, the contribution of each $N_i$ to the final baryon asymmetry is governed by the efficiency factor $k_i^f$, which can be written as:
\begin{equation}
    \begin{aligned}
    k_i^f\equiv k_i(z=\infty)=-\int_{z_{in}\to\,0}^{{z_{fin}\to\,\infty}}\dfrac{\text{d}N_i}{\text{d}z^\prime}\exp{\left[ -\sum_i\int_{z^\prime}^z W_i(z^{\prime\prime})\text{d}z^{\prime\prime}\right]}\text{d}z^\prime\,.
    \end{aligned}
    \label{eq:efficiency}
\end{equation}
We focus on the \textit{strong washout regime}, for which the Yukawa couplings are strong enough to let any species reach the equilibrium even if starting from a vanishing initial density. Therefore, the efficiency factors can be computed by inserting into the general formula \eqref{eq:efficiency} the equilibrium values $\frac{\text{d}N^{\text{eq}}_i}{\text{d}z^\prime}$ for both rate abundances. Also, the $k_i^f$ factors depend on the washout factors $K_1,K_2$, as well as on the $\delta_2=(M_2-M_1)/M_1$ factor which accounts for the mass difference between the RH neutrinos. In the left (right) upper panel of figure \ref{fig:BeyondN1DS} we plotted the $k_i^f$ factors for $\delta_2=1,\,0.1,\,0.001$ ($\delta_2=3,\,10,\,20$) values. In order to highlight the effects of the washout, we assumed $K_2\simeq 
25$. 
\begin{figure}[t]
    \centering
    \includegraphics[width=.7\textwidth]{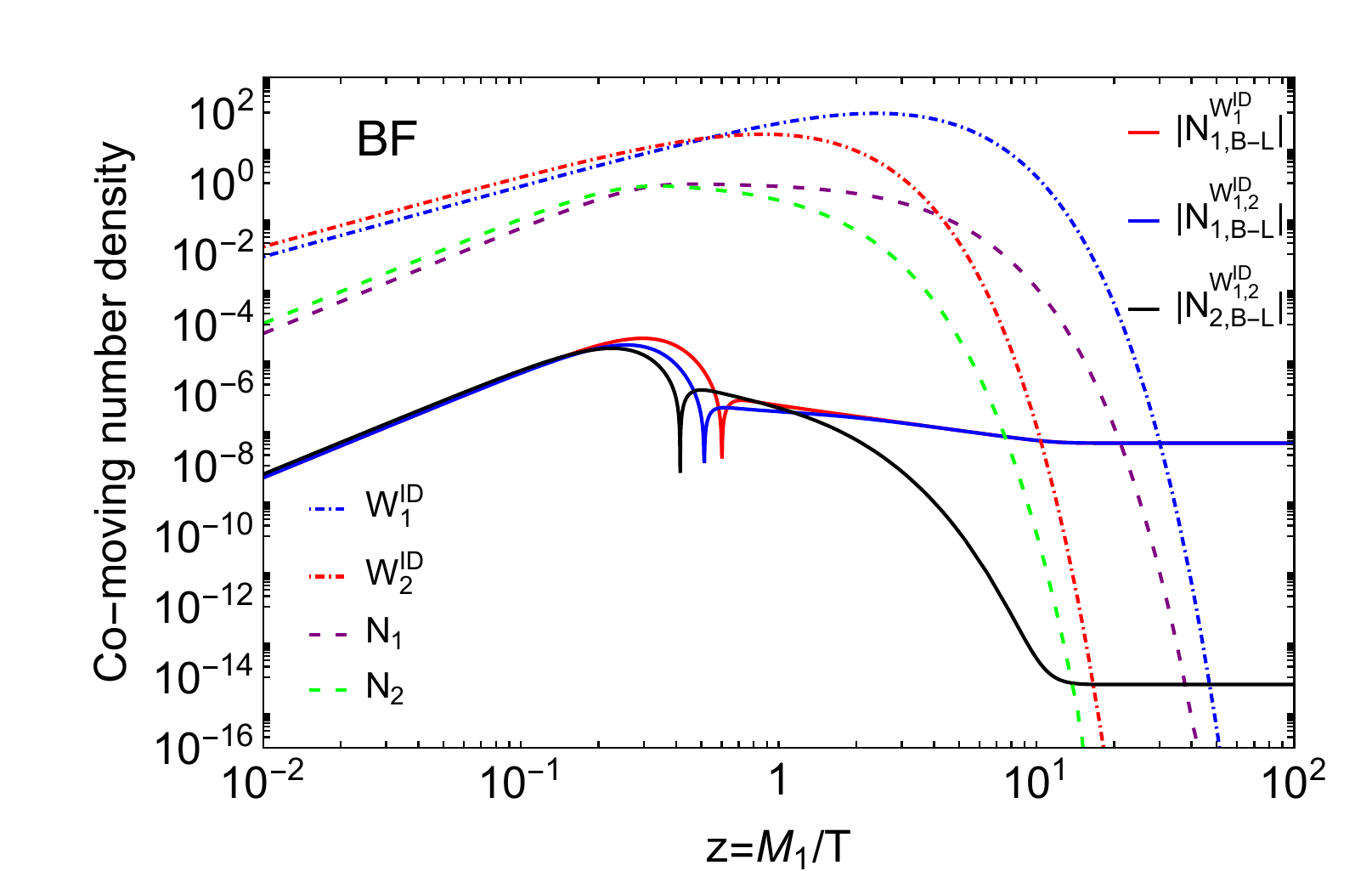}
    \caption{\textit{
$|N_{B-L}|$ as a function of $z = M_1/T$ for the best fit realization with $r=\bar{r}= 19.9$ (see section \ref{lepto_analysis} for further details). 
The colour code is the same as the bottom panels of figure \ref{fig:BeyondN1DS}. A zero initial abundance $N_i^{in}$ is assumed.}}
    \label{fig:BF}
\end{figure}\noindent
As it is evident, as the wash-out parameter increases, the hierarchy between the two efficiency factor becomes stronger, implying that the contribution to the finally BAU from the heavier neutrino can be neglected, and the $N_1-$dominated scenario is recovered.
This is confirmed by the numerical solutions of the BE in case of $\delta_2=0.3$ and $\delta_2=2$, shown in figure \ref{fig:BeyondN1DS} in the left lower panel and in the right lower panel, respectively.
Usually, a good rule of thumb to check whether the $N_1-$dominated scenario is a correct approximation consists in computing the $\delta_2$ parameter: for $\delta_2\gtrsim1.5\div5$, the contribution of the heaviest sterile neutrino $N_2$ to the leptogenesis can be safely neglected and the hierarchical N1DS is realized. Nevertheless, in a realistic model one should take into account possible correlations between the decay parameters $K_1$ and $K_2$ as well as possible hierarchies in the Yukawa couplings. Such correlations could either establish a strong hierarchy between the two efficiency factors $k_i^f$, even for relatively small values of the $\delta_2$ parameter, or limit the accessibility of the N1DS to cases where $\delta_2>5$.\\ 
In the case of our best fit realization analyzed in section \ref{lepto_analysis}, we have $r=\bar{r}=19.9$ and a value of $\delta_2\simeq 1.7$; these, along with a particular correlation between the two decay parameters $K_1\sim 10 \,K_2$, lead to a significant depletion of the asymmetry produced by $N_2$. 
Such a situation is well depicted in figure \ref{fig:BF}, where the $|B-L|$ evolution with $z$, as well as the other relevant quantities such as $N_i$ and $W_i^{ID}$ is shown. In the figure, the red and blue solid lines correspond to the asymmetry produced by $N_1$, partially depleted by the wash-out effect due to $N_1$ and $N_1+N_2$, respectively. The black line indicates the asymmetry produced by $N_2$, undergoing wash-out from $N_1+N_2$. The particular correlation between the two decay parameters together with $\delta_2\simeq 1.7$, provides a huge gap between the value at the plateau reached by the black line and the red line. 
When the strength of the inverse decays of $N_1$ reaches its maximum value, the asymmetry produced by $N_2$ is practically switched off, so it is efficiently washed-out.
In other words, the final dynamics of $|B-L|$ is governed only by the processes involving $N_1$. This is clear by analysing the evolution of the red line and the blue line: if one takes into account the asymmetry produced by $N_1$ subjected to the $N_2$ wash-out along with the $N_1$ wash-out (blue line), the final asymmetry coincides with the pure N1DS (red line).

\section{Modular forms and $S_3$ representations \label{apforms}}
The group $S_3$ is characterized by three irreducible representations: the singlet $\mathbf{1}$, the pseudo-singlet $\mathbf{1'}$ and the doublet $\mathbf{2}$. We work in a basis where the $S_3$ doublet generators are given by:
\begin{equation}
\label{genersss}
\begin{split}
\rho(S)=\frac{1}{2}\begin{pmatrix}
-1&-\sqrt{3}\\
-\sqrt{3}&1
\end{pmatrix}
\quad,\quad
\rho(T)=\begin{pmatrix}
1&0\\
0&-1
\end{pmatrix}\\ \\
(\rho(S))^2=(\rho(T))^2=(\rho(S)\rho(T))^3=\mathbbm{1}\quad\,.
\end{split}
\end{equation}

The level $N=2$ modular forms of lowest weight transform as a doublet and can be expanded as:
\begin{equation}
\label{qexp}
\begin{pmatrix}
Y_1(\tau)\\ Y_2(\tau)
\end{pmatrix}_\mathbf{2}=\begin{pmatrix}
\displaystyle\frac{7}{100}+\frac{42}{25}q+\frac{42}{25}q^2+\frac{168}{25}q^3+...\\ \\
\displaystyle\frac{14\sqrt{3}}{25}q^{1/2}(1+4q+6q^2+...)
\end{pmatrix}\,,
\end{equation}
where $q\equiv e^{2\pi i\tau}$, and $\tau=\re\tau+i\,\im\tau$. The normalization of our modular forms differs from the one chosen in \cite{Kobayashi:2018vbk}, and it is justified in \cite{Meloni:2023aru}. We stress that to our knowledge there is no physical prescription on the normalization of modular forms: it is an arbitrary choice. Recent discussions on this issue can be found in refs. \cite{deMedeirosVarzielas:2023crv, Petcov:2023fwh}. Apart from that, as discussed in the text, the ratio $|Y_2(\tau)/Y_1(\tau)|$ can be used as an expansion parameter for $\tau\in\mathcal{D}$. This is shown in figure \ref{epsmap} and the behaviour is obviously independent of the chosen normalization.

\setcounter{figure}{0}
\begin{figure}[t]
  \begin{center}
    \includegraphics[scale=0.5]{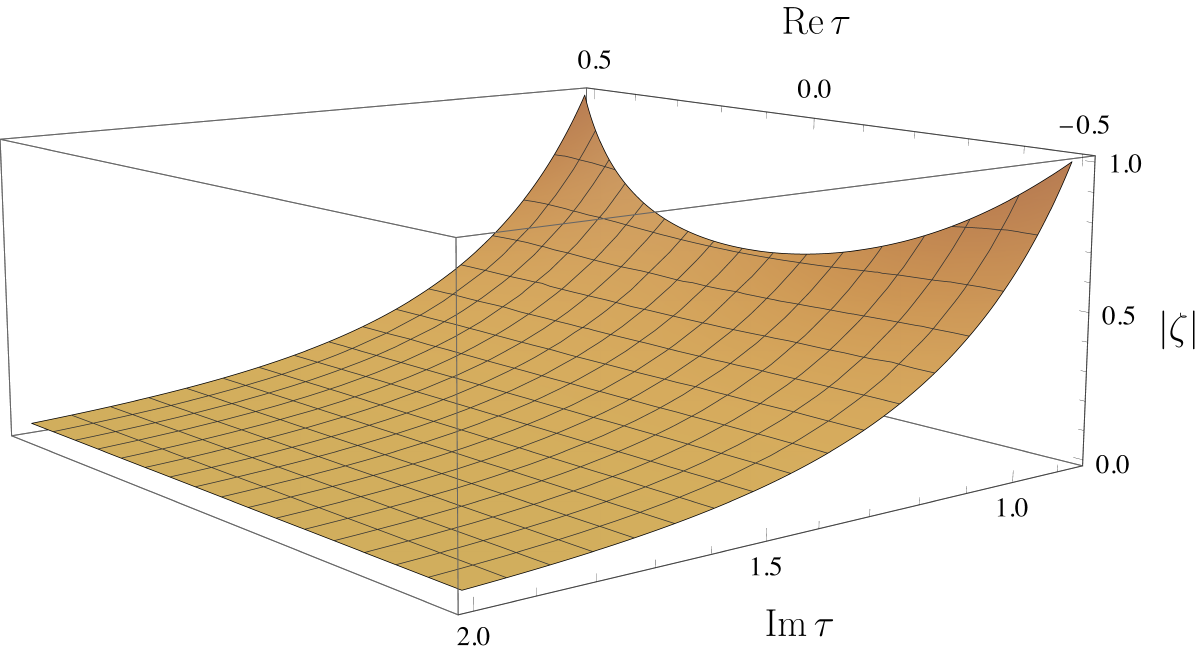}
    \caption{\small{\it Visualization of $|\zeta|=|Y_2/Y_1|$ for $\tau\in\mathcal{D}$.}}
    \label{epsmap}
  \end{center}
  \end{figure}

Labelling some $S_3$ pseudo-singlets with $y_i$ for $i=1,2,3...$, and with $\psi_{1,2},\varphi_{1,2}$ two doublet components, the tensor decomposition rules are given by:
\begin{align}
\begin{split}
\mathbf{1'}\otimes\mathbf{1'}=\mathbf{1}\quad \sim\quad  y_1y_2\,,\\
 \mathbf{1'}\otimes\mathbf{2}=\mathbf{2}\quad \sim\quad \begin{pmatrix}-y_1\psi_2 \\ y_1\psi_1\end{pmatrix}
\end{split}\,,
\end{align}

\begin{equation}
\label{cbcofff}
\mathbf{2}\otimes\mathbf{2}=\mathbf{1}\oplus\mathbf{1'}\oplus\mathbf{2}\,\begin{cases}
\mathbf{1}\quad\sim\quad \psi_1\varphi_1+\psi_2\varphi_2\\ \\
\mathbf{1'}\quad\sim\quad \psi_1\varphi_2-\psi_2\varphi_1\\ \\
\mathbf{2}\quad\sim\quad\begin{pmatrix} \psi_2\varphi_2-\psi_1\varphi_1\\ 
\psi_1\varphi_2+\psi_2\varphi_1
\end{pmatrix}\,.
\end{cases}
\end{equation}

\section{Mapping to $\mathcal{D}$ and rescaling \label{rescD}}
\setcounter{figure}{0}
\begin{figure}[t]
  \begin{center}
    \includegraphics[scale=0.35]{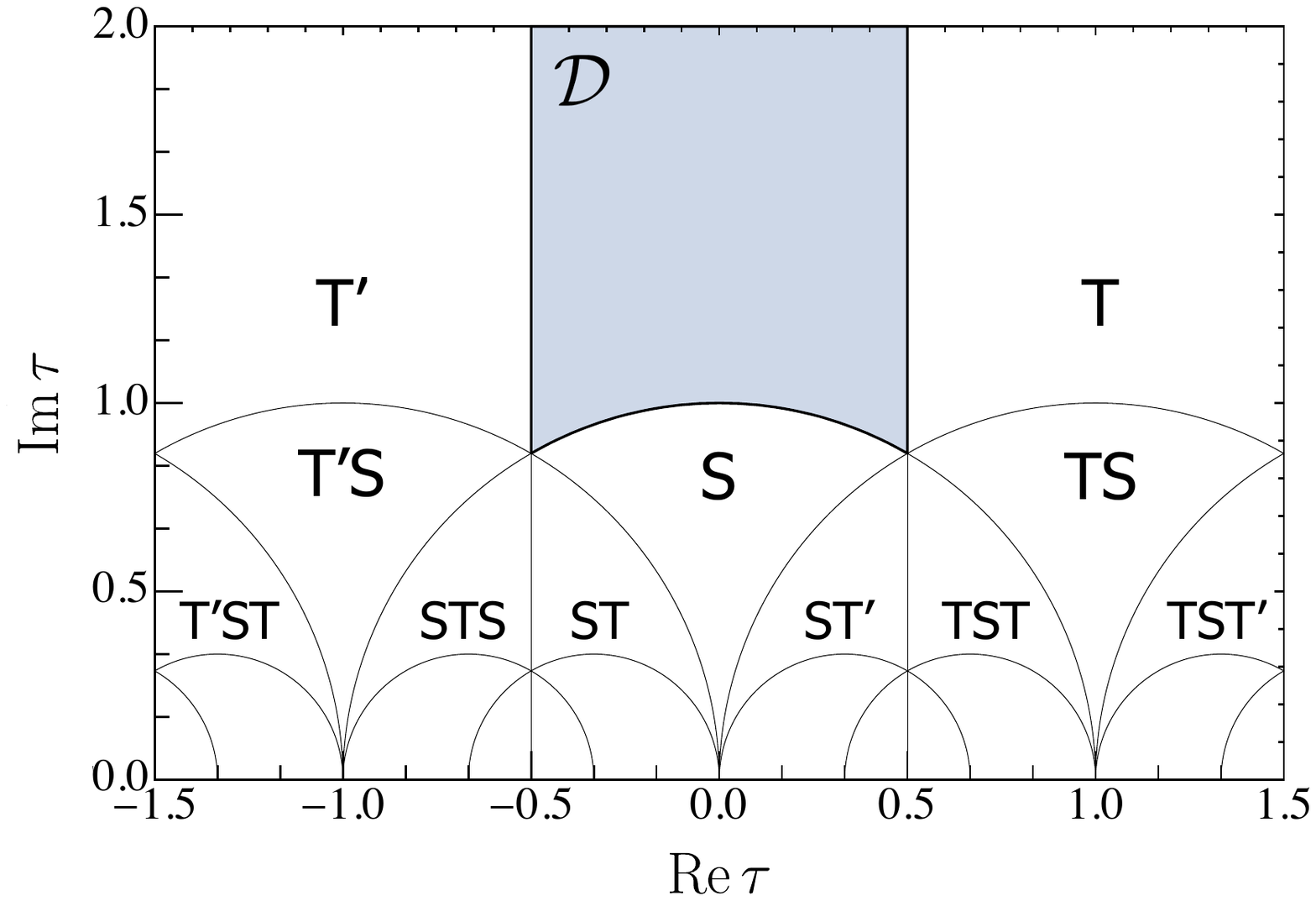}
    \caption{\small{\it Examples of regions outside the fundamental domain $\mathcal{D}$. The mapping is provided by combinations of the generators $S,T$ defined in \eqref{essti}. Here $T'\equiv T^{-1}$.}}
    \label{Dmaps}
  \end{center}
  \end{figure}
Points outside the fundamental domain \eqref{fun_domain} are redundant: some examples of modular mappings to $\mathcal{D}$ are given in figure \ref{Dmaps}. The act of mapping into $\mathcal{D}$ implies that some free parameters of the superpotential may have to be rescaled,  as argued in \cite{Novichkov:2018ovf} . The Kähler potential for the matter fields is given by:
\begin{equation}
 \sum_I(-i(\tau-\bar\tau))^{-k_I}|\varphi^{(I)}|^2=\sum_I(2\im \tau )^{-k_I}|\varphi^{(I)}|^2\,,
\end{equation}
and it is modular-invariant since under a modular transformation the relevant quantities transform as:
\begin{equation}
\label{kaltrans}
    \im\tau \quad\to \quad |c\tau+d|^{-2}\im\tau \quad \quad,\quad \varphi^{(I)}\quad \to\quad  (c\tau+d)^{-k_I}\varphi^{(I)}\,.
\end{equation}
In order to read the canonical kinetic terms from the potential, the superfields can be rescaled as:
\begin{equation}
    \varphi^{(I)}=\tilde{\varphi}^{(I)}(2\im\tau)^{k_I/2}\,,
\end{equation}
and the modular-invariant operators of the superpotential change as:
\begin{equation}
    \alpha_i \varphi^{(1)}\varphi^{(2)}\varphi^{(3)}Y_{k_Y}\quad \to \quad \alpha_i(2\im\tau)^{\frac{k_1+k_2+k_3}{2}}\tilde{\varphi}^{(1)}\tilde{\varphi}^{(2)}\tilde{\varphi}^{(3)}Y_{k_Y}\,,
    \end{equation}
where $\alpha_i$ is the associated coupling and $Y_{k_Y}$ is the required modular form of weight $k_Y$. Recall that from the modular constraints on the superpotential we know that $k_1+k_2+k_3=k_Y$. If one performed a numerical scan and extracted best-fit values for $\alpha_i$, namely $\hat{\alpha}_i$, the kinetic rescaling reads:
\begin{equation}
\label{coup_trans}
    \hat{\alpha}_i={{\alpha}_i}{(2\im\tau)^{k_Y/2}}\,.
\end{equation}
Acting with a modular transformation, $\im\tau$ transforms as in \eqref{kaltrans}. As a consequence:
\begin{equation}
    {{\alpha}_i}{(2\im\tau)^{k_Y/2}}\quad\to\quad \underbrace{{{\alpha}_i}{(2\im\tau)^{k_Y/2}}}_{\hat{\alpha_i}}|c\tau+d|^{-k_Y}\,.
\end{equation}
Thus, the coupling extracted from the numerical scan is mapped into $\hat{\alpha}_i'\equiv |c\tau+d|^{-k_Y}\hat{\alpha}_i$. By modular invariance, the parameter sets $\{\tau,\hat{\alpha}_i\}$ and $\{\gamma(\tau),\hat{\alpha}_i'\}$ lead to the same observables, as can be verified. As a common practice, the fitted parameters are the dimensionless ratios $\beta/\alpha, \gamma/\alpha, g'/g, g''/g...$ etc. These are mapped to:
\begin{equation}
\label{ratio_map}
 \frac{g_i}{g_j}\quad \to\quad \frac{g_i}{g_j} |c\tau+d|^{-k_Y^i+k_Y^j}   \,,
\end{equation}
where $k_Y^{i},k_Y^{j}$ are the weights of the modular forms associated with the $i$-th and $j$-th operators of the superpotential. In our model, all the operators from the neutrino sector have the same $k_Y^{i,j}$, thus the dimensionless ratios \eqref{ratio_map} are invariant under the modular mapping. The only non-trivial rescaling of dimensionless parameters happens in the charged-leptons sector where, in general, $k_Y^{i}\neq k_Y^{j}$ for $i\neq j$. The neutrino global mass-scale $g^2v_u^2/\Lambda$ is mapped into $|c\tau+d|^{-k_Y}g^2v_u^2/\Lambda$ where $k_Y=4$ in our case of eq. \eqref{neutrino_sup}.

\vfill
\newpage
\FloatBarrier

\renewcommand\bibname{Bibliography}
\bibliography{references}{}

\providecommand{\href}[2]{#2}\begingroup\raggedright\begin{thebibliography}{100}

\bibitem{Altarelli:2010gt}
G.~Altarelli and F.~Feruglio, \emph{{Discrete Flavor Symmetries and Models of
  Neutrino Mixing}},
  \href{https://doi.org/10.1103/RevModPhys.82.2701}{\emph{Rev. Mod. Phys.}
  {\bfseries 82} (2010) 2701}
  [\href{https://arxiv.org/abs/1002.0211}{{\ttfamily 1002.0211}}].

\bibitem{Feruglio:2019ybq}
F.~Feruglio and A.~Romanino, \emph{{Lepton flavor symmetries}},
  \href{https://doi.org/10.1103/RevModPhys.93.015007}{\emph{Rev. Mod. Phys.}
  {\bfseries 93} (2021) 015007}
  [\href{https://arxiv.org/abs/1912.06028}{{\ttfamily 1912.06028}}].

\bibitem{Feruglio:2017spp}
F.~Feruglio, \emph{{Are neutrino masses modular forms?}},  in \emph{{From My
  Vast Repertoire ...}: {Guido Altarelli's Legacy}}, A.~Levy, S.~Forte and
  G.~Ridolfi, eds., pp.~227--266 (2019),
  \href{https://doi.org/10.1142/9789813238053_0012}{DOI}
  [\href{https://arxiv.org/abs/1706.08749}{{\ttfamily 1706.08749}}].

\bibitem{Lauer:1989ax}
J.~Lauer, J.~Mas and H.P.~Nilles, \emph{{Duality and the Role of
  Nonperturbative Effects on the World Sheet}},
  \href{https://doi.org/10.1016/0370-2693(89)91190-8}{\emph{Phys. Lett. B}
  {\bfseries 226} (1989) 251}.

\bibitem{Lauer:1990tm}
J.~Lauer, J.~Mas and H.P.~Nilles, \emph{{Twisted sector representations of
  discrete background symmetries for two-dimensional orbifolds}},
  \href{https://doi.org/10.1016/0550-3213(91)90095-F}{\emph{Nucl. Phys. B}
  {\bfseries 351} (1991) 353}.

\bibitem{Ishiguro:2020tmo}
K.~Ishiguro, T.~Kobayashi and H.~Otsuka, \emph{{Landscape of Modular Symmetric
  Flavor Models}}, \href{https://doi.org/10.1007/JHEP03(2021)161}{\emph{JHEP}
  {\bfseries 03} (2021) 161}
  [\href{https://arxiv.org/abs/2011.09154}{{\ttfamily 2011.09154}}].

\bibitem{Novichkov:2022wvg}
P.P.~Novichkov, J.T.~Penedo and S.T.~Petcov, \emph{{Modular flavour symmetries
  and modulus stabilisation}},
  \href{https://doi.org/10.1007/JHEP03(2022)149}{\emph{JHEP} {\bfseries 03}
  (2022) 149} [\href{https://arxiv.org/abs/2201.02020}{{\ttfamily
  2201.02020}}].

\bibitem{Ishiguro:2022pde}
K.~Ishiguro, H.~Okada and H.~Otsuka, \emph{{Residual flavor symmetry breaking
  in the landscape of modular flavor models}},
  \href{https://doi.org/10.1007/JHEP09(2022)072}{\emph{JHEP} {\bfseries 09}
  (2022) 072} [\href{https://arxiv.org/abs/2206.04313}{{\ttfamily
  2206.04313}}].

\bibitem{Knapp-Perez:2023nty}
V.~Knapp-Perez, X.-G.~Liu, H.P.~Nilles, S.~Ramos-Sanchez and M.~Ratz,
  \emph{{Matter matters in moduli fixing and modular flavor symmetries}},
  \href{https://arxiv.org/abs/2304.14437}{{\ttfamily 2304.14437}}.

\bibitem{Kobayashi:2018vbk}
T.~Kobayashi, K.~Tanaka and T.H.~Tatsuishi, \emph{{Neutrino mixing from finite
  modular groups}},
  \href{https://doi.org/10.1103/PhysRevD.98.016004}{\emph{Phys. Rev. D}
  {\bfseries 98} (2018) 016004}
  [\href{https://arxiv.org/abs/1803.10391}{{\ttfamily 1803.10391}}].

\bibitem{Meloni:2023aru}
D.~Meloni and M.~Parriciatu, \emph{{A simplest modular S$_{3}$ model for
  leptons}}, \href{https://doi.org/10.1007/JHEP09(2023)043}{\emph{JHEP}
  {\bfseries 09} (2023) 043}
  [\href{https://arxiv.org/abs/2306.09028}{{\ttfamily 2306.09028}}].

\bibitem{Kobayashi:2018wkl}
T.~Kobayashi, Y.~Shimizu, K.~Takagi, M.~Tanimoto, T.H.~Tatsuishi and H.~Uchida,
  \emph{{Finite modular subgroups for fermion mass matrices and baryon/lepton
  number violation}},
  \href{https://doi.org/10.1016/j.physletb.2019.05.034}{\emph{Phys. Lett. B}
  {\bfseries 794} (2019) 114}
  [\href{https://arxiv.org/abs/1812.11072}{{\ttfamily 1812.11072}}].

\bibitem{Criado:2018thu}
J.C.~Criado and F.~Feruglio, \emph{{Modular Invariance Faces Precision Neutrino
  Data}}, \href{https://doi.org/10.21468/SciPostPhys.5.5.042}{\emph{SciPost
  Phys.} {\bfseries 5} (2018) 042}
  [\href{https://arxiv.org/abs/1807.01125}{{\ttfamily 1807.01125}}].

\bibitem{Kobayashi:2018scp}
T.~Kobayashi, N.~Omoto, Y.~Shimizu, K.~Takagi, M.~Tanimoto and T.H.~Tatsuishi,
  \emph{{Modular A$_{4}$ invariance and neutrino mixing}},
  \href{https://doi.org/10.1007/JHEP11(2018)196}{\emph{JHEP} {\bfseries 11}
  (2018) 196} [\href{https://arxiv.org/abs/1808.03012}{{\ttfamily
  1808.03012}}].

\bibitem{Okada:2018yrn}
H.~Okada and M.~Tanimoto, \emph{{CP violation of quarks in {$A_4$} modular
  invariance}},
  \href{https://doi.org/10.1016/j.physletb.2019.02.028}{\emph{Phys. Lett. B}
  {\bfseries 791} (2019) 54}
  [\href{https://arxiv.org/abs/1812.09677}{{\ttfamily 1812.09677}}].

\bibitem{Okada:2019uoy}
H.~Okada and M.~Tanimoto, \emph{{Towards unification of quark and lepton
  flavors in {$A_4$} modular invariance}},
  \href{https://doi.org/10.1140/epjc/s10052-021-08845-y}{\emph{Eur. Phys. J. C}
  {\bfseries 81} (2021) 52} [\href{https://arxiv.org/abs/1905.13421}{{\ttfamily
  1905.13421}}].

\bibitem{Ding:2019zxk}
G.-J.~Ding, S.F.~King and X.-G.~Liu, \emph{{Modular A$_{4}$ symmetry models of
  neutrinos and charged leptons}},
  \href{https://doi.org/10.1007/JHEP09(2019)074}{\emph{JHEP} {\bfseries 09}
  (2019) 074} [\href{https://arxiv.org/abs/1907.11714}{{\ttfamily
  1907.11714}}].

\bibitem{Kobayashi:2019xvz}
T.~Kobayashi, Y.~Shimizu, K.~Takagi, M.~Tanimoto and T.H.~Tatsuishi,
  \emph{{{$A_4$} lepton flavor model and modulus stabilization from {$S_4$}
  modular symmetry}},
  \href{https://doi.org/10.1103/PhysRevD.100.115045}{\emph{Phys. Rev. D}
  {\bfseries 100} (2019) 115045}
  [\href{https://arxiv.org/abs/1909.05139}{{\ttfamily 1909.05139}}].

\bibitem{Asaka:2019vev}
T.~Asaka, Y.~Heo, T.H.~Tatsuishi and T.~Yoshida, \emph{{Modular $A_4$
  invariance and leptogenesis}},
  \href{https://doi.org/10.1007/JHEP01(2020)144}{\emph{JHEP} {\bfseries 01}
  (2020) 144} [\href{https://arxiv.org/abs/1909.06520}{{\ttfamily
  1909.06520}}].

\bibitem{Ding:2019gof}
G.-J.~Ding, S.F.~King, X.-G.~Liu and J.-N.~Lu, \emph{{Modular S$_{4}$ and
  A$_{4}$ symmetries and their fixed points: new predictive examples of lepton
  mixing}}, \href{https://doi.org/10.1007/JHEP12(2019)030}{\emph{JHEP}
  {\bfseries 12} (2019) 030}
  [\href{https://arxiv.org/abs/1910.03460}{{\ttfamily 1910.03460}}].

\bibitem{Zhang:2019ngf}
D.~Zhang, \emph{{A modular $A_4$ symmetry realization of two-zero textures of
  the Majorana neutrino mass matrix}},
  \href{https://doi.org/10.1016/j.nuclphysb.2020.114935}{\emph{Nucl. Phys. B}
  {\bfseries 952} (2020) 114935}
  [\href{https://arxiv.org/abs/1910.07869}{{\ttfamily 1910.07869}}].

\bibitem{King:2020qaj}
S.J.D.~King and S.F.~King, \emph{{Fermion mass hierarchies from modular
  symmetry}}, \href{https://doi.org/10.1007/JHEP09(2020)043}{\emph{JHEP}
  {\bfseries 09} (2020) 043}
  [\href{https://arxiv.org/abs/2002.00969}{{\ttfamily 2002.00969}}].

\bibitem{Ding:2020yen}
G.-J.~Ding and F.~Feruglio, \emph{{Testing Moduli and Flavon Dynamics with
  Neutrino Oscillations}},
  \href{https://doi.org/10.1007/JHEP06(2020)134}{\emph{JHEP} {\bfseries 06}
  (2020) 134} [\href{https://arxiv.org/abs/2003.13448}{{\ttfamily
  2003.13448}}].

\bibitem{Asaka:2020tmo}
T.~Asaka, Y.~Heo and T.~Yoshida, \emph{{Lepton flavor model with modular $A_4$
  symmetry in large volume limit}},
  \href{https://doi.org/10.1016/j.physletb.2020.135956}{\emph{Phys. Lett. B}
  {\bfseries 811} (2020) 135956}
  [\href{https://arxiv.org/abs/2009.12120}{{\ttfamily 2009.12120}}].

\bibitem{Okada:2020brs}
H.~Okada and M.~Tanimoto, \emph{{Spontaneous CP violation by modulus $\tau$ in
  $A_4$ model of lepton flavors}},
  \href{https://doi.org/10.1007/JHEP03(2021)010}{\emph{JHEP} {\bfseries 03}
  (2021) 010} [\href{https://arxiv.org/abs/2012.01688}{{\ttfamily
  2012.01688}}].

\bibitem{Yao:2020qyy}
C.-Y.~Yao, J.-N.~Lu and G.-J.~Ding, \emph{{Modular Invariant $A_{4}$ Models for
  Quarks and Leptons with Generalized CP Symmetry}},
  \href{https://doi.org/10.1007/JHEP05(2021)102}{\emph{JHEP} {\bfseries 05}
  (2021) 102} [\href{https://arxiv.org/abs/2012.13390}{{\ttfamily
  2012.13390}}].

\bibitem{Okada:2021qdf}
H.~Okada, Y.~Shimizu, M.~Tanimoto and T.~Yoshida, \emph{{Modulus
  \ensuremath{\tau} linking leptonic CP violation to baryon asymmetry in
  A$_{4}$ modular invariant flavor model}},
  \href{https://doi.org/10.1007/JHEP07(2021)184}{\emph{JHEP} {\bfseries 07}
  (2021) 184} [\href{https://arxiv.org/abs/2105.14292}{{\ttfamily
  2105.14292}}].

\bibitem{Nomura:2021yjb}
T.~Nomura, H.~Okada and Y.~Orikasa, \emph{{Quark and lepton flavor model with
  leptoquarks in a modular {$A_4$} symmetry}},
  \href{https://doi.org/10.1140/epjc/s10052-021-09667-8}{\emph{Eur. Phys. J. C}
  {\bfseries 81} (2021) 947}
  [\href{https://arxiv.org/abs/2106.12375}{{\ttfamily 2106.12375}}].

\bibitem{Chen:2021prl}
M.-C.~Chen, V.~Knapp-Perez, M.~Ramos-Hamud, S.~Ramos-Sanchez, M.~Ratz and
  S.~Shukla, \emph{{Quasi\textendash{}eclectic modular flavor symmetries}},
  \href{https://doi.org/10.1016/j.physletb.2021.136843}{\emph{Phys. Lett. B}
  {\bfseries 824} (2022) 136843}
  [\href{https://arxiv.org/abs/2108.02240}{{\ttfamily 2108.02240}}].

\bibitem{Nomura:2022boj}
T.~Nomura, H.~Okada and Y.~Shoji, \emph{{$SU(4)_C \times SU(2)_L \times U(1)_R$
  models with modular $A_4$ symmetry}},
  \href{https://arxiv.org/abs/2206.04466}{{\ttfamily 2206.04466}}.

\bibitem{Gunji:2022xig}
Y.~Gunji, K.~Ishiwata and T.~Yoshida, \emph{{Subcritical regime of hybrid
  inflation with modular {$A_{4}$} symmetry}},
  \href{https://doi.org/10.1007/JHEP11(2022)002}{\emph{JHEP} {\bfseries 11}
  (2022) 002} [\href{https://arxiv.org/abs/2208.10086}{{\ttfamily
  2208.10086}}].

\bibitem{Devi:2023vpe}
M.R.~Devi, \emph{{Retrieving texture zeros in 3+1 active-sterile neutrino
  framework under the action of $A_4$ modular-invariants}},
  \href{https://arxiv.org/abs/2303.04900}{{\ttfamily 2303.04900}}.

\bibitem{Penedo:2018nmg}
J.T.~Penedo and S.T.~Petcov, \emph{{Lepton Masses and Mixing from Modular
  {$S_4$} Symmetry}},
  \href{https://doi.org/10.1016/j.nuclphysb.2018.12.016}{\emph{Nucl. Phys. B}
  {\bfseries 939} (2019) 292}
  [\href{https://arxiv.org/abs/1806.11040}{{\ttfamily 1806.11040}}].

\bibitem{Novichkov:2018ovf}
P.P.~Novichkov, J.T.~Penedo, S.T.~Petcov and A.V.~Titov, \emph{{Modular S$_{4}$
  models of lepton masses and mixing}},
  \href{https://doi.org/10.1007/JHEP04(2019)005}{\emph{JHEP} {\bfseries 04}
  (2019) 005} [\href{https://arxiv.org/abs/1811.04933}{{\ttfamily
  1811.04933}}].

\bibitem{deMedeirosVarzielas:2019cyj}
I.~de~Medeiros~Varzielas, S.F.~King and Y.-L.~Zhou, \emph{{Multiple modular
  symmetries as the origin of flavor}},
  \href{https://doi.org/10.1103/PhysRevD.101.055033}{\emph{Phys. Rev. D}
  {\bfseries 101} (2020) 055033}
  [\href{https://arxiv.org/abs/1906.02208}{{\ttfamily 1906.02208}}].

\bibitem{Kobayashi:2019mna}
T.~Kobayashi, Y.~Shimizu, K.~Takagi, M.~Tanimoto and T.H.~Tatsuishi, \emph{{New
  {$A_4$} lepton flavor model from {$S_4$} modular symmetry}},
  \href{https://doi.org/10.1007/JHEP02(2020)097}{\emph{JHEP} {\bfseries 02}
  (2020) 097} [\href{https://arxiv.org/abs/1907.09141}{{\ttfamily
  1907.09141}}].

\bibitem{King:2019vhv}
S.F.~King and Y.-L.~Zhou, \emph{{Trimaximal {TM$_1$} mixing with two modular
  {$S_4$} groups}},
  \href{https://doi.org/10.1103/PhysRevD.101.015001}{\emph{Phys. Rev. D}
  {\bfseries 101} (2020) 015001}
  [\href{https://arxiv.org/abs/1908.02770}{{\ttfamily 1908.02770}}].

\bibitem{Criado:2019tzk}
J.C.~Criado, F.~Feruglio and S.J.D.~King, \emph{{Modular Invariant Models of
  Lepton Masses at Levels 4 and 5}},
  \href{https://doi.org/10.1007/JHEP02(2020)001}{\emph{JHEP} {\bfseries 02}
  (2020) 001} [\href{https://arxiv.org/abs/1908.11867}{{\ttfamily
  1908.11867}}].

\bibitem{Wang:2019ovr}
X.~Wang and S.~Zhou, \emph{{The minimal seesaw model with a modular S$_{4}$
  symmetry}}, \href{https://doi.org/10.1007/JHEP05(2020)017}{\emph{JHEP}
  {\bfseries 05} (2020) 017}
  [\href{https://arxiv.org/abs/1910.09473}{{\ttfamily 1910.09473}}].

\bibitem{Wang:2020dbp}
X.~Wang, \emph{{Dirac neutrino mass models with a modular {$S_4$} symmetry}},
  \href{https://doi.org/10.1016/j.nuclphysb.2020.115247}{\emph{Nucl. Phys. B}
  {\bfseries 962} (2021) 115247}
  [\href{https://arxiv.org/abs/2007.05913}{{\ttfamily 2007.05913}}].

\bibitem{Qu:2021jdy}
B.-Y.~Qu, X.-G.~Liu, P.-T.~Chen and G.-J.~Ding, \emph{{Flavor mixing and CP
  violation from the interplay of an $S_4$ modular group and a generalized CP
  symmetry}}, \href{https://doi.org/10.1103/PhysRevD.104.076001}{\emph{Phys.
  Rev. D} {\bfseries 104} (2021) 076001}
  [\href{https://arxiv.org/abs/2106.11659}{{\ttfamily 2106.11659}}].

\bibitem{Novichkov:2018nkm}
P.P.~Novichkov, J.T.~Penedo, S.T.~Petcov and A.V.~Titov, \emph{{Modular
  {$A_{5}$} symmetry for flavour model building}},
  \href{https://doi.org/10.1007/JHEP04(2019)174}{\emph{JHEP} {\bfseries 04}
  (2019) 174} [\href{https://arxiv.org/abs/1812.02158}{{\ttfamily
  1812.02158}}].

\bibitem{Ding:2019xna}
G.-J.~Ding, S.F.~King and X.-G.~Liu, \emph{{Neutrino mass and mixing with
  {$A_5$} modular symmetry}},
  \href{https://doi.org/10.1103/PhysRevD.100.115005}{\emph{Phys. Rev. D}
  {\bfseries 100} (2019) 115005}
  [\href{https://arxiv.org/abs/1903.12588}{{\ttfamily 1903.12588}}].

\bibitem{Liu:2019khw}
X.-G.~Liu and G.-J.~Ding, \emph{{Neutrino Masses and Mixing from Double
  Covering of Finite Modular Groups}},
  \href{https://doi.org/10.1007/JHEP08(2019)134}{\emph{JHEP} {\bfseries 08}
  (2019) 134} [\href{https://arxiv.org/abs/1907.01488}{{\ttfamily
  1907.01488}}].

\bibitem{Liu:2020akv}
X.-G.~Liu, C.-Y.~Yao and G.-J.~Ding, \emph{{Modular invariant quark and lepton
  models in double covering of {$S_4$} modular group}},
  \href{https://doi.org/10.1103/PhysRevD.103.056013}{\emph{Phys. Rev. D}
  {\bfseries 103} (2021) 056013}
  [\href{https://arxiv.org/abs/2006.10722}{{\ttfamily 2006.10722}}].

\bibitem{Wang:2020lxk}
X.~Wang, B.~Yu and S.~Zhou, \emph{{Double covering of the modular {$A_5$} group
  and lepton flavor mixing in the minimal seesaw model}},
  \href{https://doi.org/10.1103/PhysRevD.103.076005}{\emph{Phys. Rev. D}
  {\bfseries 103} (2021) 076005}
  [\href{https://arxiv.org/abs/2010.10159}{{\ttfamily 2010.10159}}].

\bibitem{Yao:2020zml}
C.-Y.~Yao, X.-G.~Liu and G.-J.~Ding, \emph{{Fermion masses and mixing from the
  double cover and metaplectic cover of the {$A_5$} modular group}},
  \href{https://doi.org/10.1103/PhysRevD.103.095013}{\emph{Phys. Rev. D}
  {\bfseries 103} (2021) 095013}
  [\href{https://arxiv.org/abs/2011.03501}{{\ttfamily 2011.03501}}].

\bibitem{Novichkov:2020eep}
P.P.~Novichkov, J.T.~Penedo and S.T.~Petcov, \emph{{Double cover of modular
  {$S_4$} for flavour model building}},
  \href{https://doi.org/10.1016/j.nuclphysb.2020.115301}{\emph{Nucl. Phys. B}
  {\bfseries 963} (2021) 115301}
  [\href{https://arxiv.org/abs/2006.03058}{{\ttfamily 2006.03058}}].

\bibitem{Novichkov:2021evw}
P.P.~Novichkov, J.T.~Penedo and S.T.~Petcov, \emph{{Fermion mass hierarchies,
  large lepton mixing and residual modular symmetries}},
  \href{https://doi.org/10.1007/JHEP04(2021)206}{\emph{JHEP} {\bfseries 04}
  (2021) 206} [\href{https://arxiv.org/abs/2102.07488}{{\ttfamily
  2102.07488}}].

\bibitem{deMedeirosVarzielas:2022fbw}
I.~de~Medeiros~Varzielas, S.F.~King and M.~Levy, \emph{{Littlest modular
  seesaw}}, \href{https://doi.org/10.1007/JHEP02(2023)143}{\emph{JHEP}
  {\bfseries 02} (2023) 143}
  [\href{https://arxiv.org/abs/2211.00654}{{\ttfamily 2211.00654}}].

\bibitem{Ding:2022nzn}
G.-J.~Ding, X.-G.~Liu and C.-Y.~Yao, \emph{{A minimal modular invariant
  neutrino model}}, \href{https://doi.org/10.1007/JHEP01(2023)125}{\emph{JHEP}
  {\bfseries 01} (2023) 125}
  [\href{https://arxiv.org/abs/2211.04546}{{\ttfamily 2211.04546}}].

\bibitem{Kobayashi:2023zzc}
T.~Kobayashi and M.~Tanimoto, \emph{{Modular flavor symmetric models}},  7,
  2023 [\href{https://arxiv.org/abs/2307.03384}{{\ttfamily 2307.03384}}].

\bibitem{Ding:2023htn}
G.-J.~Ding and S.F.~King, \emph{{Neutrino Mass and Mixing with Modular
  Symmetry}},  \href{https://arxiv.org/abs/2311.09282}{{\ttfamily 2311.09282}}.

\bibitem{Fukugita:1986hr}
M.~Fukugita and T.~Yanagida, \emph{{Baryogenesis Without Grand Unification}},
  \href{https://doi.org/10.1016/0370-2693(86)91126-3}{\emph{Phys. Lett. B}
  {\bfseries 174} (1986) 45}.

\bibitem{Barbieri:1999ma}
R.~Barbieri, P.~Creminelli, A.~Strumia and N.~Tetradis, \emph{{Baryogenesis
  through leptogenesis}},
  \href{https://doi.org/10.1016/S0550-3213(00)00011-0}{\emph{Nucl. Phys. B}
  {\bfseries 575} (2000) 61}
  [\href{https://arxiv.org/abs/hep-ph/9911315}{{\ttfamily hep-ph/9911315}}].

\bibitem{Davidson_2008}
S.~Davidson, E.~Nardi and Y.~Nir, \emph{Leptogenesis},
  \href{https://doi.org/10.1016/j.physrep.2008.06.002}{\emph{Physics Reports}
  {\bfseries 466} (2008) 105–177}.

\bibitem{Buchmuller:2004nz}
W.~Buchmuller, P.~Di~Bari and M.~Plumacher, \emph{{Leptogenesis for
  pedestrians}}, \href{https://doi.org/10.1016/j.aop.2004.02.003}{\emph{Annals
  Phys.} {\bfseries 315} (2005) 305}
  [\href{https://arxiv.org/abs/hep-ph/0401240}{{\ttfamily hep-ph/0401240}}].

\bibitem{Pilaftsis:2003gt}
A.~Pilaftsis and T.E.J.~Underwood, \emph{{Resonant leptogenesis}},
  \href{https://doi.org/10.1016/j.nuclphysb.2004.05.029}{\emph{Nucl. Phys. B}
  {\bfseries 692} (2004) 303}
  [\href{https://arxiv.org/abs/hep-ph/0309342}{{\ttfamily hep-ph/0309342}}].

\bibitem{Buchmuller:2005eh}
W.~Buchmuller, R.D.~Peccei and T.~Yanagida, \emph{{Leptogenesis as the origin
  of matter}},
  \href{https://doi.org/10.1146/annurev.nucl.55.090704.151558}{\emph{Ann. Rev.
  Nucl. Part. Sci.} {\bfseries 55} (2005) 311}
  [\href{https://arxiv.org/abs/hep-ph/0502169}{{\ttfamily hep-ph/0502169}}].

\bibitem{Sakharov:1967dj}
A.D.~Sakharov, \emph{{Violation of CP Invariance, C asymmetry, and baryon
  asymmetry of the universe}},
  \href{https://doi.org/10.1070/PU1991v034n05ABEH002497}{\emph{Pisma Zh. Eksp.
  Teor. Fiz.} {\bfseries 5} (1967) 32}.

\bibitem{Gogoi:2023jzl}
J.~Gogoi, L.~Sarma and M.K.~Das, \emph{{Leptogenesis and dark matter in minimal
  inverse seesaw using $A_4$ modular symmetry}},
  \href{https://arxiv.org/abs/2311.09883}{{\ttfamily 2311.09883}}.

\bibitem{Ding:2022bzs}
G.-J.~Ding, S.F.~King, J.-N.~Lu and B.-Y.~Qu, \emph{{Leptogenesis in SO(10)
  models with A$_{4}$ modular symmetry}},
  \href{https://doi.org/10.1007/JHEP10(2022)071}{\emph{JHEP} {\bfseries 10}
  (2022) 071} [\href{https://arxiv.org/abs/2206.14675}{{\ttfamily
  2206.14675}}].

\bibitem{Kang:2022psa}
D.W.~Kang, J.~Kim, T.~Nomura and H.~Okada, \emph{{Natural mass hierarchy among
  three heavy Majorana neutrinos for resonant leptogenesis under modular
  A$_{4}$ symmetry}},
  \href{https://doi.org/10.1007/JHEP07(2022)050}{\emph{JHEP} {\bfseries 07}
  (2022) 050} [\href{https://arxiv.org/abs/2205.08269}{{\ttfamily
  2205.08269}}].

\bibitem{Mishra:2022egy}
P.~Mishra, M.K.~Behera, P.~Panda and R.~Mohanta, \emph{{Type III seesaw under
  $A_4$ modular symmetry with leptogenesis}},
  \href{https://doi.org/10.1140/epjc/s10052-022-11074-6}{\emph{Eur. Phys. J. C}
  {\bfseries 82} (2022) 1115}
  [\href{https://arxiv.org/abs/2204.08338}{{\ttfamily 2204.08338}}].

\bibitem{Dasgupta:2021ggp}
A.~Dasgupta, T.~Nomura, H.~Okada, O.~Popov and M.~Tanimoto, \emph{{Dirac
  Radiative Neutrino Mass with Modular Symmetry and Leptogenesis}},
  \href{https://arxiv.org/abs/2111.06898}{{\ttfamily 2111.06898}}.

\bibitem{Kashav:2021zir}
M.~Kashav and S.~Verma, \emph{{Broken scaling neutrino mass matrix and
  leptogenesis based on A$_{4}$ modular invariance}},
  \href{https://doi.org/10.1007/JHEP09(2021)100}{\emph{JHEP} {\bfseries 09}
  (2021) 100} [\href{https://arxiv.org/abs/2103.07207}{{\ttfamily
  2103.07207}}].

\bibitem{Behera:2020sfe}
M.K.~Behera, S.~Mishra, S.~Singirala and R.~Mohanta, \emph{{Implications of A4
  modular symmetry on neutrino mass, mixing and leptogenesis with linear
  seesaw}}, \href{https://doi.org/10.1016/j.dark.2022.101027}{\emph{Phys. Dark
  Univ.} {\bfseries 36} (2022) 101027}
  [\href{https://arxiv.org/abs/2007.00545}{{\ttfamily 2007.00545}}].

\bibitem{Behera:2022wco}
M.K.~Behera and R.~Mohanta, \emph{{Linear Seesaw in $A_5'$ Modular Symmetry
  With Leptogenesis}},
  \href{https://doi.org/10.3389/fphy.2022.854595}{\emph{Front. in Phys.}
  {\bfseries 10} (2022) 854595}
  [\href{https://arxiv.org/abs/2201.10429}{{\ttfamily 2201.10429}}].

\bibitem{Novichkov:2019sqv}
P.P.~Novichkov, J.T.~Penedo, S.T.~Petcov and A.V.~Titov, \emph{{Generalised
  {CP} Symmetry in Modular-Invariant Models of Flavour}},
  \href{https://doi.org/10.1007/JHEP07(2019)165}{\emph{JHEP} {\bfseries 07}
  (2019) 165} [\href{https://arxiv.org/abs/1905.11970}{{\ttfamily
  1905.11970}}].

\bibitem{King:1999mb}
S.F.~King, \emph{{Large mixing angle MSW and atmospheric neutrinos from single
  right-handed neutrino dominance and U(1) family symmetry}},
  \href{https://doi.org/10.1016/S0550-3213(00)00109-7}{\emph{Nucl. Phys. B}
  {\bfseries 576} (2000) 85}
  [\href{https://arxiv.org/abs/hep-ph/9912492}{{\ttfamily hep-ph/9912492}}].

\bibitem{Frampton:2002qc}
P.H.~Frampton, S.L.~Glashow and T.~Yanagida, \emph{{Cosmological sign of
  neutrino CP violation}},
  \href{https://doi.org/10.1016/S0370-2693(02)02853-8}{\emph{Phys. Lett. B}
  {\bfseries 548} (2002) 119}
  [\href{https://arxiv.org/abs/hep-ph/0208157}{{\ttfamily hep-ph/0208157}}].

\bibitem{King:2002qh}
S.F.~King, \emph{{Leptogenesis MNS link in unified models with natural neutrino
  mass hierarchy}},
  \href{https://doi.org/10.1103/PhysRevD.67.113010}{\emph{Phys. Rev. D}
  {\bfseries 67} (2003) 113010}
  [\href{https://arxiv.org/abs/hep-ph/0211228}{{\ttfamily hep-ph/0211228}}].

\bibitem{Raidal:2002xf}
M.~Raidal and A.~Strumia, \emph{{Predictions of the most minimal seesaw
  model}}, \href{https://doi.org/10.1016/S0370-2693(02)03124-6}{\emph{Phys.
  Lett. B} {\bfseries 553} (2003) 72}
  [\href{https://arxiv.org/abs/hep-ph/0210021}{{\ttfamily hep-ph/0210021}}].

\bibitem{King:2002nf}
S.F.~King, \emph{{Constructing the large mixing angle MNS matrix in seesaw
  models with right-handed neutrino dominance}},
  \href{https://doi.org/10.1088/1126-6708/2002/09/011}{\emph{JHEP} {\bfseries
  09} (2002) 011} [\href{https://arxiv.org/abs/hep-ph/0204360}{{\ttfamily
  hep-ph/0204360}}].

\bibitem{Xing:2020ald}
Z.-z.~Xing and Z.-h.~Zhao, \emph{{The minimal seesaw and leptogenesis models}},
  \href{https://doi.org/10.1088/1361-6633/abf086}{\emph{Rept. Prog. Phys.}
  {\bfseries 84} (2021) 066201}
  [\href{https://arxiv.org/abs/2008.12090}{{\ttfamily 2008.12090}}].

\bibitem{Esteban:2020cvm}
I.~Esteban, M.C.~Gonzalez-Garcia, M.~Maltoni, T.~Schwetz and A.~Zhou,
  \emph{{The fate of hints: updated global analysis of three-flavor neutrino
  oscillations}}, \href{https://doi.org/10.1007/JHEP09(2020)178}{\emph{JHEP}
  {\bfseries 09} (2020) 178}
  [\href{https://arxiv.org/abs/2007.14792}{{\ttfamily 2007.14792}}].

\bibitem{Super-Kamiokande:2017yvm}
{\scshape Super-Kamiokande} collaboration, \emph{{Atmospheric neutrino
  oscillation analysis with external constraints in Super-Kamiokande I-IV}},
  \href{https://doi.org/10.1103/PhysRevD.97.072001}{\emph{Phys. Rev. D}
  {\bfseries 97} (2018) 072001}
  [\href{https://arxiv.org/abs/1710.09126}{{\ttfamily 1710.09126}}].

\bibitem{Antusch:2003kp}
S.~Antusch, J.~Kersten, M.~Lindner and M.~Ratz, \emph{{Running neutrino masses,
  mixings and CP phases: Analytical results and phenomenological
  consequences}},
  \href{https://doi.org/10.1016/j.nuclphysb.2003.09.050}{\emph{Nucl. Phys. B}
  {\bfseries 674} (2003) 401}
  [\href{https://arxiv.org/abs/hep-ph/0305273}{{\ttfamily hep-ph/0305273}}].

\bibitem{Maki:1962mu}
Z.~Maki, M.~Nakagawa and S.~Sakata, \emph{{Remarks on the unified model of
  elementary particles}}, \href{https://doi.org/10.1143/PTP.28.870}{\emph{Prog.
  Theor. Phys.} {\bfseries 28} (1962) 870}.

\bibitem{Pontecorvo:1957qd}
B.~Pontecorvo, \emph{{Inverse beta processes and nonconservation of lepton
  charge}}, {\emph{Zh. Eksp. Teor. Fiz.} {\bfseries 34} (1957) 247}.

\bibitem{Novichkov:2021cgl}
P.~Novichkov, \emph{{Aspects of the Modular Symmetry Approach to Lepton
  Flavour}}, Ph.D. thesis, SISSA, Trieste, 2021.

\bibitem{Altarelli:2010at}
G.~Altarelli and G.~Blankenburg, \emph{{Different $SO(10)$ Paths to Fermion
  Masses and Mixings}},
  \href{https://doi.org/10.1007/JHEP03(2011)133}{\emph{JHEP} {\bfseries 03}
  (2011) 133} [\href{https://arxiv.org/abs/1012.2697}{{\ttfamily 1012.2697}}].

\bibitem{eBOSS:2020yzd}
{\scshape eBOSS} collaboration, \emph{{Completed SDSS-IV extended Baryon
  Oscillation Spectroscopic Survey: Cosmological implications from two decades
  of spectroscopic surveys at the Apache Point Observatory}},
  \href{https://doi.org/10.1103/PhysRevD.103.083533}{\emph{Phys. Rev. D}
  {\bfseries 103} (2021) 083533}
  [\href{https://arxiv.org/abs/2007.08991}{{\ttfamily 2007.08991}}].

\bibitem{GIUDICE200489}
G.~Giudice, A.~Notari, M.~Raidal, A.~Riotto and A.~Strumia, \emph{Towards a
  complete theory of thermal leptogenesis in the sm and mssm},
  \href{https://doi.org/https://doi.org/10.1016/j.nuclphysb.2004.02.019}{\emph{Nuclear
  Physics B} {\bfseries 685} (2004) 89}.

\bibitem{tHooft:1976snw}
G.~'t~Hooft, \emph{{Computation of the Quantum Effects Due to a
  Four-Dimensional Pseudoparticle}},
  \href{https://doi.org/10.1103/PhysRevD.14.3432}{\emph{Phys. Rev. D}
  {\bfseries 14} (1976) 3432}.

\bibitem{Davidson:2002qv}
S.~Davidson and A.~Ibarra, \emph{{A Lower bound on the right-handed neutrino
  mass from leptogenesis}},
  \href{https://doi.org/10.1016/S0370-2693(02)01735-5}{\emph{Phys. Lett. B}
  {\bfseries 535} (2002) 25}
  [\href{https://arxiv.org/abs/hep-ph/0202239}{{\ttfamily hep-ph/0202239}}].

\bibitem{Pradler:2006qh}
J.~Pradler and F.D.~Steffen, \emph{{Thermal gravitino production and collider
  tests of leptogenesis}},
  \href{https://doi.org/10.1103/PhysRevD.75.023509}{\emph{Phys. Rev. D}
  {\bfseries 75} (2007) 023509}
  [\href{https://arxiv.org/abs/hep-ph/0608344}{{\ttfamily hep-ph/0608344}}].

\bibitem{Samanta_2020}
R.~Samanta and M.~Sen, \emph{Flavoured leptogenesis and cp {$\mu-\tau$}
  symmetry}, \href{https://doi.org/10.1007/jhep01(2020)193}{\emph{Journal of
  High Energy Physics} {\bfseries 2020} (2020) }.

\bibitem{Blanchet_2012}
S.~Blanchet and P.D.~Bari, \emph{The minimal scenario of leptogenesis},
  \href{https://doi.org/10.1088/1367-2630/14/12/125012}{\emph{New Journal of
  Physics} {\bfseries 14} (2012) 125012}.

\bibitem{Blanchet_2006}
S.~Blanchet and P.D.~Bari, \emph{Leptogenesis beyond the limit of hierarchical
  heavy neutrino masses},
  \href{https://doi.org/10.1088/1475-7516/2006/06/023}{\emph{Journal of
  Cosmology and Astroparticle Physics} {\bfseries 2006} (2006) 023}.

\bibitem{Fong:2012buy}
C.S.~Fong, E.~Nardi and A.~Riotto, \emph{{Leptogenesis in the Universe}},
  \href{https://doi.org/10.1155/2012/158303}{\emph{Adv. High Energy Phys.}
  {\bfseries 2012} (2012) 158303}
  [\href{https://arxiv.org/abs/1301.3062}{{\ttfamily 1301.3062}}].

\bibitem{2020}
{\scshape Planck} collaboration, \emph{{Planck 2018 results. VI. Cosmological
  parameters}},
  \href{https://doi.org/10.1051/0004-6361/201833910}{\emph{Astron. Astrophys.}
  {\bfseries 641} (2020) A6}
  [\href{https://arxiv.org/abs/1807.06209}{{\ttfamily 1807.06209}}].

\bibitem{Buchm_ller_2004}
W.~Buchmüller, P.D.~Bari and M.~Plümacher, \emph{Some aspects of thermal
  leptogenesis}, \href{https://doi.org/10.1088/1367-2630/6/1/105}{\emph{New
  Journal of Physics} {\bfseries 6} (2004) 105–105}.

\bibitem{Endoh_2004}
T.~Endoh, T.~Morozumi and Z.~Xiong, \emph{Primordial lepton family asymmetries
  in seesaw model}, \href{https://doi.org/10.1143/ptp.111.123}{\emph{Progress
  of Theoretical Physics} {\bfseries 111} (2004) 123–149}.

\bibitem{Abada_2006}
A.~Abada, S.~Davidson, F.-X.~Josse-Michaux, M.~Losada and A.~Riotto,
  \emph{Flavour issues in leptogenesis},
  \href{https://doi.org/10.1088/1475-7516/2006/04/004}{\emph{Journal of
  Cosmology and Astroparticle Physics} {\bfseries 2006} (2006) 004–004}.

\bibitem{Nardi_2006}
E.~Nardi, Y.~Nir, E.~Roulet and J.~Racker, \emph{The importance of flavor in
  leptogenesis},
  \href{https://doi.org/10.1088/1126-6708/2006/01/164}{\emph{Journal of High
  Energy Physics} {\bfseries 2006} (2006) 164–164}.

\bibitem{Campbell_1992}
B.A.~Campbell, S.~Davidson, J.~Ellis and K.A.~Olive, \emph{On the baryon,
  lepton-flavour and right-handed electron asymmetries of the universe},
  \href{https://doi.org/10.1016/0370-2693(92)91079-o}{\emph{Physics Letters B}
  {\bfseries 297} (1992) 118–124}.

\bibitem{Cline_1994}
J.M.~Cline, K.~Kainulainen and K.A.~Olive, \emph{Protecting the primordial
  baryon asymmetry from erasure by sphalerons},
  \href{https://doi.org/10.1103/physrevd.49.6394}{\emph{Physical Review D}
  {\bfseries 49} (1994) 6394–6409}.

\bibitem{Covi_1996}
L.~Covi, E.~Roulet and F.~Vissani, \emph{Cp violating decays in leptogenesis
  scenarios}, \href{https://doi.org/10.1016/0370-2693(96)00817-9}{\emph{Physics
  Letters B} {\bfseries 384} (1996) 169–174}.

\bibitem{Hamaguchi_2002}
K.~Hamaguchi, H.~Murayama and T.~Yanagida, \emph{Leptogenesis from a
  n1-dominated early universe},
  \href{https://doi.org/10.1103/physrevd.65.043512}{\emph{Physical Review D}
  {\bfseries 65} (2002) }.

\bibitem{Pascoli_2007}
S.~Pascoli, S.~Petcov and A.~Riotto, \emph{Leptogenesis and low energy
  cp-violation in neutrino physics},
  \href{https://doi.org/10.1016/j.nuclphysb.2007.02.019}{\emph{Nuclear Physics
  B} {\bfseries 774} (2007) 1–52}.

\bibitem{Branco_2009}
G.C.~Branco, R.~Gonzalez~Felipe, M.N.~Rebelo and H.~Serodio, \emph{{Resonant
  leptogenesis and tribimaximal leptonic mixing with A(4) symmetry}},
  \href{https://doi.org/10.1103/PhysRevD.79.093008}{\emph{Phys. Rev. D}
  {\bfseries 79} (2009) 093008}
  [\href{https://arxiv.org/abs/0904.3076}{{\ttfamily 0904.3076}}].

\bibitem{Buchm_ller_1996}
W.~Buchmüller and M.~Plümacher, \emph{Baryon asymmetry and neutrino mixing},
  \href{https://doi.org/10.1016/s0370-2693(96)01232-4}{\emph{Physics Letters B}
  {\bfseries 389} (1996) 73–77}.

\bibitem{Pramanick:2024gvu}
R.~Pramanick, T.S.~Ray and A.~Sil, \emph{{Towards a more complete description
  of hybrid leptogenesis}},  \href{https://arxiv.org/abs/2401.12189}{{\ttfamily
  2401.12189}}.

\bibitem{Buchm_ller_2003}
W.~Buchmüller, P.~Di~Bari and M.~Plümacher, \emph{The neutrino mass window
  for baryogenesis},
  \href{https://doi.org/10.1016/s0550-3213(03)00449-8}{\emph{Nuclear Physics B}
  {\bfseries 665} (2003) 445–468}.

\bibitem{Buchmuller:2002rq}
W.~Buchmuller, P.~Di~Bari and M.~Plumacher, \emph{{Cosmic microwave background,
  matter - antimatter asymmetry and neutrino masses}},
  \href{https://doi.org/10.1016/S0550-3213(02)00737-X}{\emph{Nucl. Phys. B}
  {\bfseries 643} (2002) 367}
  [\href{https://arxiv.org/abs/hep-ph/0205349}{{\ttfamily hep-ph/0205349}}].

\bibitem{croon2019stability}
D.~Croon, N.~Fernandez, D.~McKeen and G.~White, \emph{Stability, reheating and
  leptogenesis},  2019.

\bibitem{Giudice_1999}
G.F.~Giudice, A.~Riotto, I.~Tkachev and M.~Peloso, \emph{Production of massive
  fermions at preheating and leptogenesis},
  \href{https://doi.org/10.1088/1126-6708/1999/08/014}{\emph{Journal of High
  Energy Physics} {\bfseries 1999} (1999) 014–014}.

\bibitem{Khlopov:1984pf}
M.Y.~Khlopov and A.D.~Linde, \emph{{Is It Easy to Save the Gravitino?}},
  \href{https://doi.org/10.1016/0370-2693(84)91656-3}{\emph{Phys. Lett. B}
  {\bfseries 138} (1984) 265}.

\bibitem{Alexander:2006lty}
S.H.-S.~Alexander, M.E.~Peskin and M.M.~Sheikh-Jabbari,
  \emph{{Gravi-Leptogenesis: Leptogenesis from Gravity Waves in Pseudo-scalar
  Driven Inflation Models}}, {\emph{eConf} {\bfseries C0605151} (2006) 0022}
  [\href{https://arxiv.org/abs/hep-ph/0701139}{{\ttfamily hep-ph/0701139}}].

\bibitem{Adshead:2017znw}
P.~Adshead, A.J.~Long and E.I.~Sfakianakis, \emph{{Gravitational Leptogenesis,
  Reheating, and Models of Neutrino Mass}},
  \href{https://doi.org/10.1103/PhysRevD.97.043511}{\emph{Phys. Rev. D}
  {\bfseries 97} (2018) 043511}
  [\href{https://arxiv.org/abs/1711.04800}{{\ttfamily 1711.04800}}].

\bibitem{Bertuzzo_2011}
E.~Bertuzzo, P.~Di~Bari and
  L.~Marzola\href{https://doi.org/10.1016/j.nuclphysb.2011.03.027}{\emph{Nuclear
  Physics B} {\bfseries 849} (2011) 521–548}.

\bibitem{Di_Bari_2013}
P.~Di~Bari and
  L.~Marzola\href{https://doi.org/10.1016/j.nuclphysb.2013.10.027}{\emph{Nuclear
  Physics B} {\bfseries 877} (2013) 719–751}.

\bibitem{Chianese_2018}
M.~Chianese and P.~Di~Bari, \emph{Strong thermal so(10)-inspired leptogenesis
  in the light of recent results from long-baseline neutrino experiments},
  \href{https://doi.org/10.1007/jhep05(2018)073}{\emph{Journal of High Energy
  Physics} {\bfseries 2018} (2018) }.

\bibitem{PhysRevD.42.3344}
J.A.~Harvey and M.S.~Turner, \emph{Cosmological baryon and lepton number in the
  presence of electroweak fermion-number violation},
  \href{https://doi.org/10.1103/PhysRevD.42.3344}{\emph{Phys. Rev. D}
  {\bfseries 42} (1990) 3344}.

\bibitem{Dev:2017wwc}
B.~Dev, M.~Garny, J.~Klaric, P.~Millington and D.~Teresi, \emph{{Resonant
  enhancement in leptogenesis}},
  \href{https://doi.org/10.1142/S0217751X18420034}{\emph{Int. J. Mod. Phys. A}
  {\bfseries 33} (2018) 1842003}
  [\href{https://arxiv.org/abs/1711.02863}{{\ttfamily 1711.02863}}].

\bibitem{Pilaftsis_2005}
A.~Pilaftsis and T.E.J.~Underwood, \emph{Electroweak-scale resonant
  leptogenesis},
  \href{https://doi.org/10.1103/physrevd.72.113001}{\emph{Physical Review D}
  {\bfseries 72} (2005) }.

\bibitem{Xing_2007}
Z.-z.~Xing and S.~Zhou, \emph{Tri-bimaximal neutrino mixing and
  flavor-dependent resonant leptogenesis},
  \href{https://doi.org/10.1016/j.physletb.2007.08.009}{\emph{Physics Letters
  B} {\bfseries 653} (2007) 278–287}.

\bibitem{Arcadi_2023}
G.~Arcadi, S.~Marciano and D.~Meloni, \emph{Neutrino mixing and leptogenesis in
  a $l_e-l_\mu -l_\tau $ model},
  \href{https://doi.org/10.1140/epjc/s10052-023-11268-6}{\emph{The European
  Physical Journal C} {\bfseries 83} (2023) }.

\bibitem{Jung_2022}
K.-Y.~Jung and K.~Siyeon, \emph{Light sterile neutrino and leptogenesis},
  \href{https://doi.org/10.1007/s40042-022-00674-w}{\emph{Journal of the Korean
  Physical Society} {\bfseries 81} (2022) 1211–1224}.

\bibitem{Blanchet:2008pw}
S.~Blanchet and P.~Di~Bari, \emph{{New aspects of leptogenesis bounds}},
  \href{https://doi.org/10.1016/j.nuclphysb.2008.08.026}{\emph{Nucl. Phys. B}
  {\bfseries 807} (2009) 155}
  [\href{https://arxiv.org/abs/0807.0743}{{\ttfamily 0807.0743}}].

\bibitem{DUNE:2020jqi}
{\scshape DUNE} collaboration, \emph{{Long-baseline neutrino oscillation
  physics potential of the DUNE experiment}},
  \href{https://doi.org/10.1140/epjc/s10052-020-08456-z}{\emph{Eur. Phys. J. C}
  {\bfseries 80} (2020) 978}
  [\href{https://arxiv.org/abs/2006.16043}{{\ttfamily 2006.16043}}].

\bibitem{Hyper-KamiokandeProto-:2015xww}
{\scshape Hyper-Kamiokande Proto-} collaboration, \emph{{Physics potential of a
  long-baseline neutrino oscillation experiment using a J-PARC neutrino beam
  and Hyper-Kamiokande}},
  \href{https://doi.org/10.1093/ptep/ptv061}{\emph{PTEP} {\bfseries 2015}
  (2015) 053C02} [\href{https://arxiv.org/abs/1502.05199}{{\ttfamily
  1502.05199}}].

\bibitem{JUNO:2015zny}
{\scshape JUNO} collaboration, \emph{{Neutrino Physics with JUNO}},
  \href{https://doi.org/10.1088/0954-3899/43/3/030401}{\emph{J. Phys. G}
  {\bfseries 43} (2016) 030401}
  [\href{https://arxiv.org/abs/1507.05613}{{\ttfamily 1507.05613}}].

\bibitem{IceCube-Gen2:2019fet}
{\scshape IceCube-Gen2} collaboration, \emph{{Combined sensitivity to the
  neutrino mass ordering with JUNO, the IceCube Upgrade, and PINGU}},
  \href{https://doi.org/10.1103/PhysRevD.101.032006}{\emph{Phys. Rev. D}
  {\bfseries 101} (2020) 032006}
  [\href{https://arxiv.org/abs/1911.06745}{{\ttfamily 1911.06745}}].

\bibitem{KamLAND-Zen:2022tow}
{\scshape KamLAND-Zen} collaboration, \emph{{Search for the Majorana Nature of
  Neutrinos in the Inverted Mass Ordering Region with KamLAND-Zen}},
  \href{https://doi.org/10.1103/PhysRevLett.130.051801}{\emph{Phys. Rev. Lett.}
  {\bfseries 130} (2023) 051801}
  [\href{https://arxiv.org/abs/2203.02139}{{\ttfamily 2203.02139}}].

\bibitem{KATRIN:2021uub}
{\scshape KATRIN} collaboration, \emph{{Direct neutrino-mass measurement with
  sub-electronvolt sensitivity}},
  \href{https://doi.org/10.1038/s41567-021-01463-1}{\emph{Nature Phys.}
  {\bfseries 18} (2022) 160}
  [\href{https://arxiv.org/abs/2105.08533}{{\ttfamily 2105.08533}}].

\bibitem{Kolb:1990vq}
E.W.~Kolb and M.S.~Turner, \emph{{The Early Universe}}, vol.~69 (1990),
  \href{https://doi.org/10.1201/9780429492860}{10.1201/9780429492860}.

\bibitem{deMedeirosVarzielas:2023crv}
I.~de~Medeiros~Varzielas, M.~Levy, J.T.~Penedo and S.T.~Petcov, \emph{{Quarks
  at the modular $S_4$ cusp}},
  \href{https://arxiv.org/abs/2307.14410}{{\ttfamily 2307.14410}}.

\bibitem{Petcov:2023fwh}
S.T.~Petcov, \emph{{On the Normalisation of the Modular Forms in Modular
  Invariant Theories of Flavour}},
  \href{https://arxiv.org/abs/2311.04185}{{\ttfamily 2311.04185}}.

\end{thebibliography}\endgroup
\bibliographystyle{JHEP}

\end{document}